\documentclass[12pt]{article}
\usepackage[english]{babel}
\usepackage[utf8]{inputenc}
\usepackage[T1]{fontenc}
\usepackage{amsbsy}
\usepackage{amsfonts}
\usepackage{amsmath}
\usepackage{amssymb}
\usepackage{amsthm}
\usepackage{graphicx} 
\usepackage[pdftex]{color}
\usepackage{enumerate}
\usepackage{float}
\usepackage{geometry, calc, color}
\usepackage{indentfirst}
\usepackage{longtable}
\usepackage{multirow}
\usepackage{natbib}
\usepackage{xr-hyper}
\usepackage{hyperref}
\usepackage{caption}
\usepackage{subcaption}
\usepackage{booktabs}
\usepackage{xcolor}
\usepackage[linesnumbered,ruled,vlined]{algorithm2e}

\newtheorem{theorem}{Theorem}[section]
\newtheorem{definition}{Definition}[section]

\newtheorem{proposition}[theorem]{Proposition}

\newtheorem{remark}[theorem]{Remark}

\DeclareMathOperator*{\plim}{plim}

\hypersetup{
  colorlinks=true,
  linkcolor=red,
  citecolor=teal,
  filecolor=blue,
  urlcolor=teal,
}
\newcommand{\blind}{1}

\SetCommentSty{mycommfont}

\begin{document}

\def\spacingset#1{\renewcommand{\baselinestretch}%
	{#1}\small\normalsize} \spacingset{1}

\if1\blind
{
	\title{\bf Bounded GARCH Processes} 
	 \author{Wagner Barreto-Souza\footnote{Email: \textcolor{teal}{\texttt{wagner.barreto-souza@ucd.ie}}}
	 \hspace{.2cm}\\
	 {\normalsize \it School of Mathematics and Statistics, University College Dublin, Belfield, Republic of Ireland}}
	\maketitle
} \fi
\if0\blind
{
	\bigskip
	\bigskip
	\bigskip
	\begin{center}
		{\LARGE\bf Bounded GARCH Processes}
	\end{center}
	\medskip
} \fi

\bigskip
\addtocontents{toc}{\protect\setcounter{tocdepth}{1}}

\begin{abstract}
We propose a novel class of Bounded GARCH (B-GARCH) processes designed to characterise the conditional heteroskedasticity of continuous time series constrained within a compact interval. While standard GARCH specifications are tailored for $\mathbb{R}$-valued data, many empirical series, such as realised correlation, unemployment rates, expenditure shares, and relative humidity, are bounded. Fitting a traditional GARCH model to such data creates a severe support mismatch, frequently generating predictions and density forecasts that breach the natural boundaries of the data ({\it boundary violation}). Furthermore, conventional approaches utilising non-linear link transformations onto the real line might induce interpretability issues and estimation biases. To address these limitations, our proposed B-GARCH framework ensures a proper predictive distribution and an equation for the dynamic conditional variance with a parameter constraint that strictly preserves the bounded support of the entire process almost surely. We establish the fundamental probabilistic properties of the process, proving strict stationarity and ergodicity, and address parameter estimation via conditional maximum likelihood. The asymptotic consistency and normality of the estimators are rigorously derived, and their finite-sample performance is validated through Monte Carlo simulations. An empirical application to the monthly realised correlation between the Vanguard Real Estate ETF (VNQ) and the iShares 20+ Year Treasury Bond ETF (TLT) demonstrates that B-GARCH models effectively capture volatility clustering while completely eliminating {\it boundary violation} issue that plague traditional GARCH models.
\end{abstract}



\section{Introduction}

Generalised Autoregressive Conditional Heteroscedastic (GARCH) processes \citep{eng1982, bol1986} serve as the benchmark framework for modelling volatility in finance and econometrics. Over the decades, several classical extensions to the GARCH framework have been proposed, including the persistence-focused integrated GARCH (IGARCH) model \citep{engbol1986}, the asymmetric exponential GARCH (EGARCH) model \citep{nel1991}, and the threshold-based GJR-GARCH model \citep{gjr1993}. Furthermore, the asymmetric power ARCH (APARCH) formulation \citep{dineta1993} added flexibility through power transformations, while long-memory dynamics were subsequently captured via the fractionally integrated GARCH (FIGARCH) model \citep{bolmik1996}. A great account on this extensive field is provided by \cite{frazak2019}. The domain remains highly vibrant, with recent advancements including the matrix GARCH model by \cite{yuetal2025}, the multivariate realised GARCH framework of \cite{archaketal2026} for high-dimensional tracking, the $\mathbb{Z}$-valued smooth transition GARCH model by \cite{zhuetal2026}, alongside the graph-theoretic copula-GARCH model advanced by \cite{chanetal2026} to capture dynamic conditional dependence, just to name a few.

Although standard GARCH specifications are inherently designed for unconstrained, $\mathbb{R}$-valued processes, many economic and physical indicators are strictly confined by mathematical or structural boundaries. Prominent examples include realised correlation, unemployment rates, budget expenditure shares, political approval ratings, market indices, and relative humidity, as well as returns of bounded time series. A key methodological challenge when modelling bounded time series is ensuring that the conditional variance framework structurally respects its physical boundaries without sacrificing empirical flexibility. While traditional unconstrained formulations require parameter restrictions solely to maintain long-term stability, their unrestricted intercept means the total impact of the model's components can easily exceed the natural range. Under a standard Gaussian or similar assumption, prediction issues might be faced. Due to the $\mathbb R$ support, out-of-sample density forecasts, confidence intervals, and extreme-value predictions routinely assign positive probability to impossible regions that lie completely outside the natural range of the data, which we will refer to here as {\it boundary violation}.

The necessity of developing specialised methodologies for bounded time series has been emphasised by \cite{cav2005} and \cite{cavxu2014}, who provided the first rigorous attempts to address unit root testing in this context. Their work demonstrated that standard tests developed for $\mathbb{R}$-valued series often fail when applied to bounded data, emphasizing that accounting for boundary constraints is both effective and necessary. The relevance on further developments of methods for handling bounded time series  has been also highlighted by \cite{gra2010} in his latest paper, as subsequently noted by \cite{cavxu2014}. A common strategy for tackling boundedness is to transform the data onto the real line, for instance, via a logit transformation, and subsequently apply classical methods such as ARIMA or GARCH models. This approach, however, possesses several drawbacks. As argued by \cite{gorkoo2023}, such transformations induce non-linear effects and often result in a loss of interpretability. Furthermore, retransforming estimated quantities such as predictions back to the original scale might introduce significant biases, and there is no guarantee that such predictions or associated confidence intervals will remain within the natural range of the data.

To avoid these issues, various alternatives based on the beta distribution have been explored. For instance, \cite{roccri2009} proposed a class of beta ARMA models using a maximum likelihood approach, while \cite{casetal2012} developed a Bayesian framework for beta autoregressive models. Further contributions utilising beta-based formulations include \cite{dasetal2011}, \cite{guovar2014}, \cite{pumetal2021}, \cite{gorkoo2023}, \cite{graetal2025}, and \cite{caaetal2026}; additionally, a semiparametric approach was introduced by \cite{akndim2023}. A common feature shared by all these models is that they are designed to model the conditional mean, thereby modelling the conditional variance only indirectly. Despite this growing body of literature, no attempt has yet been made to primarily model the conditional volatility of bounded time series.

The objective of this paper is to fill this gap. We propose a novel class of Bounded GARCH (B-GARCH) models that retain the essential features of the classic GARCH framework while respecting boundary constraints. Specifically, for a time series $\{Y_t\}_{t\in\mathbb{N}}$ and the filtration $\mathcal{F}_{t-1}$ generated by its past, our model satisfies $E(Y_t|\mathcal{F}_{t-1})=0$, while the conditional variance $h_t \equiv \text{Var}(Y_t|\mathcal{F}_{t-1})$ follows a stochastic dynamic driven by lagged squared observations and previous conditional variances. Crucially, the process preserves the support $Y_t \in (-1, 1)$, allowing us to model bounded series without the need for non-linear transformations. This framework is naturally suited for $(-1, 1)$-valued data, such as realised correlation or returns of bounded time series, or more generally, it can be easily adapted for $(a, b)$-valued series via a simple linear transformation mapping $(a,b)$ onto $(-1,1)$. Beyond introducing the model specification, this paper establishes the rigorous theoretical foundations necessary for reliable statistical inference. We characterise the fundamental stochastic properties of the B-GARCH processes, by establishing their strict stationarity and ergodicity, and develop a conditional maximum likelihood estimation framework. It is worth noting that the proof of strict stationarity and ergodicity for our processes cannot be established using standard theory (top Lyapunov exponent theory) from classical GARCH models, so our proof represents a novel contribution to the literature (see Remark \ref{rem:non_standard_proof} for details). We also prove the consistency and asymptotic normality of the proposed estimators, thereby providing the essential asymptotic theory required for inference. Our Monte Carlo simulations validate our theoretical results and demonstrate robust finite-sample uncertainty quantification.

We propose a generalisation of the B-GARCH processes to accommodate asymmetry, which is an important feature observed in bounded time series. To capture this, we incorporate a dynamic conditional mean rather than restricting the process to a zero-mean specification. We emphasise that the bounded GARCH framework prevents the aforementioned {\it boundary violation} through a straightforward two-part design: it combines a naturally bounded conditional distribution with a proper parameter constraint. An empirical illustration on realised asset correlations demonstrates the superior predictive performance of bounded GARCH models while highlighting the boundary violations produced by the classical GARCH specification. While demonstrated here on realised correlations, the proposed methodology provides a general statistical framework for continuous processes on compact domains across the physical, economic, and social sciences, with immediate applications to climatology (relative humidity), labour economics (unemployment rates), public finance (budget expenditure shares), and political science (approval ratings), just to name a few.

The remainder of this paper is organised as follows. Section \ref{sec:model_def} introduces the B-GARCH class and explores its mathematical properties. Parameter estimation and associated asymptotics are detailed in Section \ref{sec:estimation}. Section \ref{sec:MC} reports the results of our Monte Carlo study. We generalise the B-GARCH models to accomodate asymmetry in Section \ref{sec:generalisations}. We apply the bounded GARCH processes to analyse the monthly realised correlation between the Vanguard Real Estate ETF (VNQ) and the iShares 20+ Year Treasury Bond ETF (TLT) from January 1, 2010, to December 31, 2025 in Section \ref{sec:app}, benchmarking their performance against the classical GARCH model. Finally, Section \ref{sec:conclusion} provides concluding remarks and suggests avenues for future research. Technical proofs are presented in the Appendix.

\section{Model specification and properties}\label{sec:model_def}

To establish the formal framework for the bounded GARCH class, we first construct a symmetric, bounded innovation distribution parameterised directly by its variance. Let $X$ denote a random variable following a classic symmetric beta distribution on the support $(0,1)$, characterised by the probability density function
\begin{equation*}
	f(x)=\dfrac{\Gamma(\phi)}{[\Gamma(\phi/2)]^2}[x(1-x)]^{\phi/2-1},\quad 0<x<1,	
\end{equation*}	
where $\Gamma(\cdot)$ represents the gamma function and $\phi>0$ is a precision parameter. The mean and variance of $X$ are $\mathbb{E}(X)=1/2$ and $\mbox{Var}(X)=\dfrac{1}{4(1+\phi)}$, respectively. 
To map this support to the symmetric domain $(-1,1)$ suitable for bounded returns and indicators, we use the linear transformation $Y=2X-1$. To render the distribution structurally amenable to volatility modelling, we introduce the reparameterisation $\phi=\dfrac{1}{\sigma^2}-1$, which implicitly restricts the variance parameter to the unit interval, $\sigma^2\in(0,1)$. By a straightforward change of variables, the density of the transformed variable $Y$ initially assumes the form
\begin{equation*}
	f(y)=\dfrac{4\Gamma(1/\sigma^2-1)}{2^{1/\sigma^2}\left[\Gamma\left(\frac{1/\sigma^2-1}{2}\right)\right]^2}(1-y^2)^{(1/\sigma^2-3)/2},\quad -1<y<1.	
\end{equation*}	
Evaluating this expression via the Legendre duplication formula $\Gamma(z)\Gamma(z+1/2)=2^{1-2z}\sqrt{\pi}\Gamma(2z)$ with $z=\dfrac{1/\sigma^2-1}{2}$, yields the density function of $Y$ in the form
\begin{equation}\label{density}
	f(y)=\dfrac{\Gamma\left(\frac{1}{2\sigma^2}\right)}{\sqrt{\pi}\Gamma\left(\frac{1/\sigma^2-1}{2}\right)}(1-y^2)^{(1/\sigma^2-3)/2},\quad -1<y<1.	
\end{equation}	
Under this specific parameterisation, we have that $\mathbb{E}(Y)=0$, $\mbox{Var}(Y)=\sigma^2$, $\mathbb{E}(Y^3)=0$, and $\mathbb{E}(Y^4)=\dfrac{3(\sigma^2)^2}{2\sigma^2+1}$. Hereafter, we denote a random variable with density \eqref{density} as $Y\sim\operatorname{Beta}(0,\sigma^2)$. In what follows, we define our processes to model the volatility of bounded time series and explore some of their properties. 

\begin{definition}\label{def:modeldel}[$\operatorname{B-GARCH}(p,q)$ processes]
	We say that a stochastic sequence $\{Y_t\}_{t\in \mathbb{Z}}$ is a bounded $\operatorname{GARCH}(p,q)$ process if 
	\begin{equation*}
		Y_t \mid \mathcal{F}_{t-1} \sim \operatorname{Beta}(0,h_t), \quad \forall t \geq 1,
	\end{equation*}
	where $\mathcal{F}_{t-1}=\sigma(Y_{t-1},h_{t-1},Y_{t-2},h_{t-2},\ldots)$ is the filtration generated by the past history, and the conditional variance $h_t=\mbox{Var}(Y_t\mid \mathcal{F}_{t-1})$ evolves according to
	\begin{equation}\label{ht:condvariance}
		h_t = \alpha_0 + \sum_{i=1}^p \alpha_i Y^2_{t-i} + \sum_{j=1}^q \beta_j h_{t-j}, \quad \forall t \geq 1,
	\end{equation}	
	with parameters satisfying $\alpha_0>0$, $\alpha_i \geq 0$ for $i=1,\ldots,p$, $\beta_j \geq 0$ for $j=1,\ldots,q$, and the strict upper bound constraint 
	\begin{equation}\label{const}
		\alpha_0 + \sum_{i=1}^p \alpha_i + \sum_{j=1}^q \beta_j < 1.
	\end{equation}
\end{definition}

\begin{remark}\label{rem:beta_choice}
The choice of the beta distribution in Definition~\ref{def:modeldel} is motivated by key theoretical and empirical considerations. Just as the Gaussian distribution is the standard benchmark for unconstrained real-valued data, the beta distribution serves as the natural counterpart for continuous data restricted to a compact interval, both within time series econometrics \citep[see][]{roccri2009,gorkoo2023} and across broader statistical methodology.
	
	Theoretically, the $\text{Beta}(0, h_t)$ family is essential for establishing strict stationarity and ergodicity of the B-GARCH processes; see proof of Theorem \ref{thm:staterg} and Remark \ref{rem:non_standard_proof}. From an estimation standpoint, the beta log-density offers analytical advantages in terms of a tractable likelihood evaluation, score derivations, and an analytical conditional information matrix. In particular, this enables us to show that the score sequence forms an exact martingale difference sequence (Theorem \ref{thm:mds}), facilitating a clean proof of consistency and asymptotic normality for the conditional maximum likelihood estimator (Theorem \ref{thm:AN}).
	
	From an empirical perspective, the $\text{Beta}(0, h_t)$ distribution adapts its shape naturally as the conditional variance $h_t$ evolves. The conditional density transitions smoothly from a bell-shaped curve under low volatility ($h_t < 1/3$) to a uniform density ($h_t = 1/3$), and to a U-shaped density under high volatility ($h_t > 1/3$), placing more probability mass near the boundaries during turbulent periods. 
\end{remark}

\begin{remark}\label{rem:comparison}
	For a time series bounded within the interval $(-1,1)$, the associated conditional (and marginal) variance is structurally restricted to the interval $(0,1)$. The constraint specified in \eqref{const} mathematically guarantees that this natural boundary is strictly enforced almost surely. 
	
	In contrast, a traditional GARCH model under a Gaussian or Student-$t$ specification dictates that while weak stationarity requires $\sum_{i=1}^p \alpha_i + \sum_{j=1}^q \beta_j < 1$, the unconstrained intercept $\alpha_0 > 0$ implies that the total sum $\alpha_0 + \sum_{i=1}^p \alpha_i + \sum_{j=1}^q \beta_j$ can easily exceed 1. Consequently, fitting a traditional GARCH model to bounded data frequently yields estimated conditional variances and out-of-sample density predictions that violate the physical boundaries of the support. 
	One might argue that a traditional GARCH specification could simply be estimated under the constraint (\ref{const}); however, while this restricts the conditional variance parameter, the $\mathbb R$ support of a Gaussian or similar distribution ensures that the model will still generate out-of-sample observations and density forecasts, so still suffering from boundary violation. Our proposed framework inherently overcomes this flaw. Moreover, modelling the time series on its natural domain avoids non-linear data transformations (such as logit), which distort volatility persistence parameters and introduce estimation bias when transforming predictions back to the original scale.
\end{remark}

Let  $\{Y_t\}_{t\geq1}$ be a $\mbox{B-GARCH}(p,q)$ process. Under stationarity, it follows that 
$$E(Y_t)=E(E(Y_t|\mathcal F_{t-1}))=0$$ 
and 
\begin{eqnarray*}
&&\mbox{Var}(Y_t)=E(\mbox{Var}(Y_t|\mathcal F_{t-1}))=E(h_t)=\alpha_0+\displaystyle\sum_{i=1}^p\alpha_iE(Y^2_{t-i})+\displaystyle\sum_{j=1}^q\beta_jE(h_{t-j}) \implies\\
&& \mbox{Var}(Y_t)=\dfrac{\alpha_0}{1-\displaystyle\sum_{i=1}^p\alpha_i-\displaystyle\sum_{j=1}^q\beta_j},
\end{eqnarray*}	
where we have used that $E(Y^2_{t-i})=E(h_{t-j})=\mbox{Var}(Y_t)$ for all $i$ and $j$ due to the stationarity assumption.

\begin{figure}[ht!]
	\centering
	\includegraphics[width=1\linewidth]{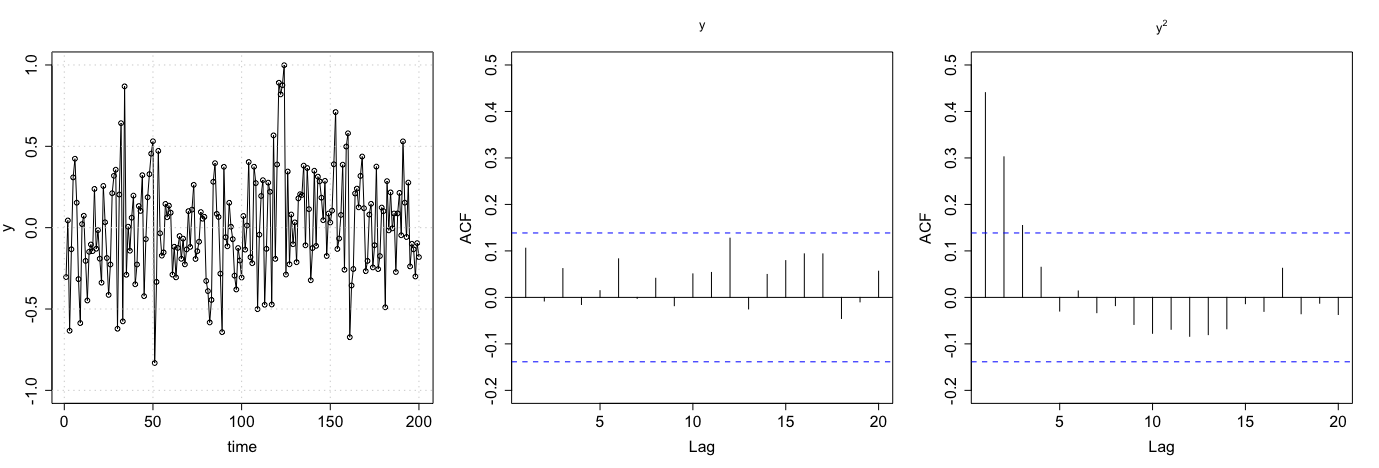}
\caption{Simulated trajectory of $200$ observations from a $\operatorname{B-GARCH}(1,1)$ model with parameters $\alpha_0=0.05$, $\alpha_1=0.4$, and $\beta_1=0.2$ (left panel), alongside the sample autocorrelation functions (ACFs) of $y_t$ (middle panel) and $y_t^2$ (right panel).}
\label{fig:sim_traj}
\end{figure}

For visual inspection, a simulated trajectory $y_1,\ldots,y_{200}$ from a $\operatorname{B-GARCH}(1,1)$ model with parameters $\alpha_0=0.05$, $\alpha_1=0.4$, and $\beta_1=0.2$ is presented in Figure~\ref{fig:sim_traj}, along with the empirical autocorrelations of $y_t$ and $y_t^2$. The ACF plots reveal that while $y_t$ mirrors an uncorrelated white noise sequence, $y_t^2$ exhibits significant serial depconsieendence. This dichotomy aligns perfectly with the stylistic facts of traditional Gaussian GARCH processes, demonstrating that the bounded framework preserves core volatility clustering properties.

To evaluate the tail behaviour of the proposed models, we address the kurtosis properties of the $\operatorname{B-GARCH}(1,0)$ process ($h_t = \alpha_0 + \alpha_1 Y_{t-1}^2$). Using the notation $\operatorname{Kurt}(Y_t)$ for the kurtosis of $Y_t$, the next proposition details a set of inequalities that bound this metric, followed by a discussion on their interpretation.

\begin{proposition}\label{prop:kurtosis}
	Let $\{Y_t\}_{t\geq 1}\sim\mbox{B-GARCH}(1,0)$. Then,\\
	(i) the kurtosis of the Beta noise (case $\alpha_1=0$) is $\kappa_{noise}\equiv \mbox{Kurt}(Y_t)=\dfrac{3}{2\alpha_0+1}$; \\
	(ii) $\mbox{Kurt}(Y_t)\geq\dfrac{3(1-\alpha_1)(1-\alpha_1^2)}{2\alpha_0+(1-\alpha_1)(1-3\alpha_1^2)}$;\\
	(iii) $\mbox{Kurt}(Y_t)\leq \kappa_{noise}$ for $\alpha_0\geq\dfrac{1-\alpha_1}{2}$.
\end{proposition}

\begin{remark}
	Due to the non-linear form of the fourth conditional moment in terms of $h_t$, that is $E(Y_t^4|\mathcal F_{t-1})=\dfrac{3h_t^2}{2h_t+1}$, we are not able to obtain an explicit form for the kurtosis of $Y_t$ in contrast with the classic GARCH model. Still, we provide some inequalities that can be of interest including an inferior limit. An interesting result is (iii), where it is shown that the kurtosis for the B-GARCH model is less than that from the beta noise process. One can initially think that this goes on the other direction to what is expected since the classic GARCH has heavy-tails than the normal noise. One important feature is that $Y_t$ is bounded and kurtosis is a measure to compare tails with a normal distribution. The noise in our case is $\mbox{Beta}(0,\alpha_0)$ distributed and so the normal reference does not make sense here. The kurtosis for our model needs to be carefully interpreted. For example, the  $\mbox{Beta}(0,\alpha_0)$ distribution has a bell-shaped density function for $\alpha_0<1/3$, while the marginal density of $Y_t$ (stationary distribution for $\alpha_1>0$) could be $U$-shaped. This phenomenon is illustrated in Figure~\ref{fig:kurt}, which displays the $\operatorname{Beta}(0,1/4)$ and standard normal density functions alongside the empirical distribution (histogram) of the stationary distribution obtained via simulation with $\alpha_1=0.4$. We also present a case where the noise density function is $U$-shaped displayed on the right of Figure \ref{fig:kurt} with $\alpha_0=0.5$ and $\alpha_1=0.3$. In this case, both noise and stationary densities are $U$-shaped, and again the stationary density has a heavier tail when compared to the $\mbox{Beta}(0,\alpha_0)$ noise. These explain why the Beta noise has a higher kurtosis compared to the B-GARCH(1,0) model, but, in practice, the tails of the $\mbox{Beta}(0,\alpha_0)$ density are lighter than those of the B-GARCH(1,0) model. All in all, result (iii) is in line with what is expected.
	\begin{figure}[ht!]
		\centering
		\includegraphics[width = .5\linewidth]{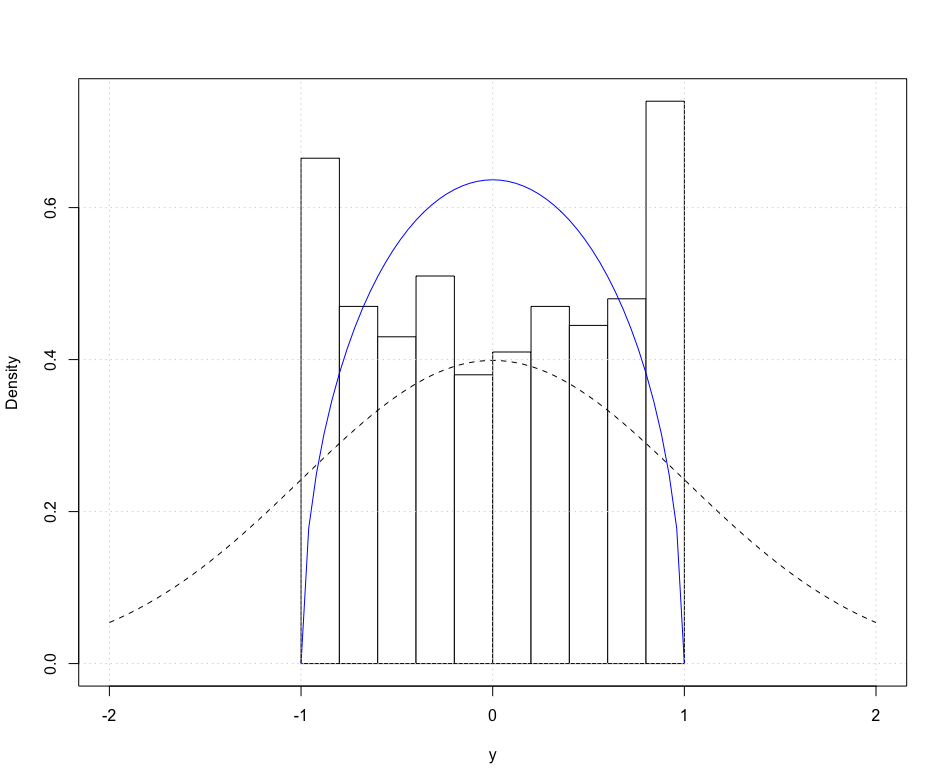}\includegraphics[width = .5\linewidth]{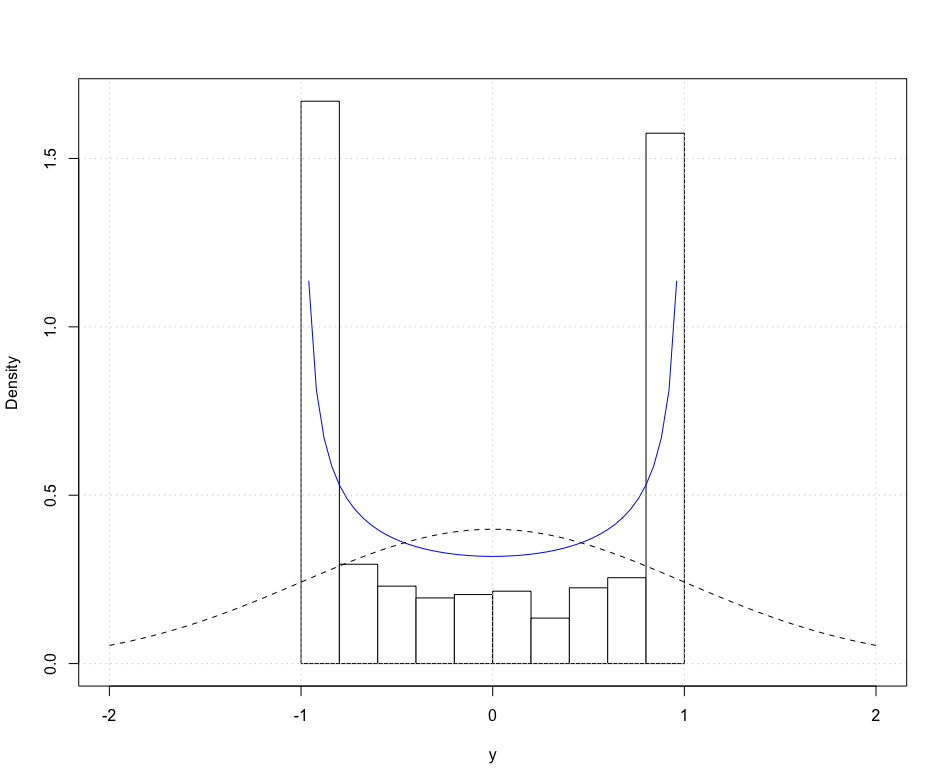}
		\caption{$\mbox{Beta}(0,\alpha_0)$ (solid curve) and standard normal (dashed curve) density functions and the empirical stationary distribution (histogram) of the B-GARCH(1,0) process with $(\alpha_0,\alpha_1)=(0.25,0.4)$ (on the left) and $(\alpha_0,\alpha_1)=(0.5,0.3)$ (on the right).}
		\label{fig:kurt}
	\end{figure}
\end{remark}

We now address core theoretical properties of the B-GARCH processes. For GARCH-type models, the main interest relies on the squared process $Y_t^2$ rather than $Y_t$ since $Y_t^2$ carries the information on the volatility. Further, as we will see in the section on estimation of parameters, the likelihood function depends on $\{(Y_t^2,h_t)\}_{t\geq1}$. So, in order to establish asymptotic results for the maximum likelihood estimators, we need to prove desirable stochastic properties of the above bivariate process, which are provided in the following theorem. For simplicity of notation, we establish the result for the first-order case $(p,q)=(1,1)$, though the underlying properties extend naturally to the general $\mbox{B-GARCH}(p,q)$ specification.

\begin{theorem}\label{thm:staterg}
Let  $\{Y_t\}_{t\geq1}$	be a B-GARCH(1,1) process and define $Z_t=Y_t^2$. Then,\\
(i) there exists a unique stationary solution to $\{(Z_t,h_t)\}_{t\geq1}$, with distribution, say $\Pi$;\\
(ii) the process $\{(Z_t,h_t)\}_{t\geq1}$ initiated with $(Z_0,h_0)\sim\Pi$ is strictly stationary and ergodic.\\
\end{theorem}	

\begin{remark}
	\label{rem:non_standard_proof}
Standard proofs of strict stationarity and ergodicity for classical GARCH processes (for instance, see \cite{frazak2019}) rely fundamentally on the multiplicative noise factorisation $Y_t = \sqrt{h_t}\eta_t$, where $\eta_t \stackrel{i.i.d.}{\sim} N(0,1)$, converting the volatility dynamic into a linear random-coefficient autoregression amenable to top Lyapunov exponent analysis. For the B-GARCH framework, $Y_t \sim \text{Beta}(0, h_t)$ cannot be factorised into a scale factor and an independent innovation, rendering classical linear techniques inapplicable. To resolve this structural limitation, our proof (see Appendix) establishes stationarity and ergodicity using Markov chain contraction theory \citep{douwin2008}. The core technical innovation relies on the strict concavity of the beta log-density ratio, and the application of stochastic order results \citep{yu2009} to establish a tight $L_1$ bound on expected quantile differences, providing the necessary stochastic contraction to guarantee stationarity and ergodicity.
\end{remark}

\begin{remark}
	Assume that $\{(Z_t,h_t)\}_{t\geq1}$ is initiated at $(Z_0,h_0)\sim\pi$, with $\pi$ being the invariant measure, where $\{Y_t\}_{t\geq1}\sim\mbox{B-GARCH}(1,1)$ and $Z_t=Y_t^2$. Theorem \ref{thm:staterg} then establishes the joint strictly stationarity and ergodicity of $\{(Z_t,h_t)\}_{t\geq1}$. In particular, this implies the marginal strictly stationarity and ergodicity of $\{h_t\}_{t\geq1}$, which is enough to conclude that such properties are also valid for  $\{Y_t\}_{t\geq1}$ (under the above initialisation).
\end{remark}	

\section{Estimation of parameters and asymptotics}\label{sec:estimation}

This section is devoted to conditional maximum likelihood estimation and its asymptotic properties. Let $y_1,\ldots,y_n$ be a trajectory from a $\mbox{B-GARCH}(p,q)$ process, $\boldsymbol\theta=(\alpha_0,\alpha_1,\ldots,\alpha_p,\beta_1,\ldots,\beta_q)^\top$ be the parameter vector, $\Theta=\big\{\boldsymbol\theta: \alpha_0>0, \alpha_1,\ldots,\alpha_p\geq0, \beta_1,\ldots,\beta_q\geq0, \alpha_0+\sum_{i=1}^p\alpha_i+\sum_{j=1}^q\beta_j<1\big\}$ be the parameter space, and denote $m=\max(p,q)$. The conditional log-likelihood function (given $y_1,\ldots,y_m$) is
\begin{eqnarray*}
\ell(\boldsymbol\theta)=\sum_{t=m+1}^n\ell_t(\boldsymbol\theta)=\sum_{t=m+1}^n\log f(y_t|\mathcal F_{t-1}),
\end{eqnarray*}	
where 
\begin{eqnarray*}
\ell_t(\boldsymbol\theta)&=&\log f(y_t|\mathcal F_{t-1})\\
&=&(1/h_t-3)\log(1-y_t^2)/2+\log\Gamma\left(\frac{1}{2h_t}\right)-\log\Gamma\left(\frac{1/h_t-1}{2}\right)-\dfrac{1}{2}\log\pi\\
&\propto&\dfrac{\log(1-y_t^2)}{2h_t}+\log\Gamma\left(\frac{1}{2h_t}\right)-\log\Gamma\left(\frac{1/h_t-1}{2}\right),
\end{eqnarray*}	
for $t=m+1,\ldots,n$. The score function is ${\bf U}(\boldsymbol\theta)=\partial\ell(\boldsymbol\theta)/\partial\boldsymbol\theta=\sum_{t=m+1}^n{\bf U}_t(\boldsymbol\theta)$, with
\begin{eqnarray}\label{eq:Utk}
	{\bf U}_t(\boldsymbol\theta)\equiv\left\{-\log(1-y_t^2)-\psi\left(\frac{1}{2h_t}\right)+\psi\left(\frac{1/h_t-1}{2}\right)\right\}\dfrac{1}{2h^2_t}
	\dfrac{\partial h_t}{\partial\boldsymbol\theta},
\end{eqnarray}
where $\psi(x)=d\log\Gamma(x)/dx$ is the digamma function. The conditional maximum likelihood estimator (CMLE) of $\boldsymbol{\theta}$ is $\widehat{\boldsymbol{\theta}}=\mbox{argmax}_{\theta}\ell(\boldsymbol\theta)$.

By using results from the proof of the Theorem \ref{thm:mds} (see Appendix), we obtain that the conditional expectation of the terms of second-order derivatives of the conditional log-likelihood function assumes a simple form. With this, the standard errors of the conditional maximum likelihood estimates can be consistently estimated via the information matrix ${\bf I}(\boldsymbol\theta)$ given by
\begin{eqnarray}\label{information}
	{\bf I}(\boldsymbol\theta)=\sum_{t=m+1}^nE\left(-\dfrac{\partial^2\ell_t(\boldsymbol\theta)}{\partial\boldsymbol\theta\partial\boldsymbol\theta^\top}\Big|\mathcal F_{t-1}\right)=\sum_{t=m+1}^n\left\{\psi'\left(\dfrac{1/h_t-1}{2}\right)-\psi'\left(\dfrac{1}{2h_t}\right)\right\}\dfrac{1}{4h_t^4}\dfrac{\partial h_t}{\partial\boldsymbol\theta}{\dfrac{\partial h_t}{\partial\boldsymbol\theta}}^\top,
\end{eqnarray}	
where $\psi'(x)=d\psi(x)/dx$ is the trigamma function. The efficiency of the information matrix to assess standard errors will be studied in finite-sample sizes in Section \ref{sec:MC}.

In what follows, we will consider the B-GARCH(1,1) model. The next theorem establishes that the terms involved in the score function given in (\ref{eq:Utk}) form a Martingale Difference Sequence (MDS), which will be helpful to prove the asymptotic normality of the CMLE $\widehat{\boldsymbol{\theta}}$.

\begin{theorem}\label{thm:mds}
Assume that $\{Y_t\}_{t\geq1}\sim\mbox{B-GARCH}(1,1)$ and let ${\bf U}_t(\boldsymbol\theta)$ for $t\geq2$ as defined in (\ref{eq:Utk}). The sequence $\{{\bf U}_t(\boldsymbol\theta)\}_{t\geq2}$ is a MDS with respect to the filtration $\{\mathcal F_t\}_{t\geq1}$. Further, ${\bf U}_t(\boldsymbol\theta)$ is square-integrable for all $t\geq2$.
\end{theorem}

The existence, uniqueness, consistency and asymptotic normality of the CMLE $\widehat{\boldsymbol{\theta}}$ are established in the next theorem.

\begin{theorem}\label{thm:AN}
	Assume that $Y_1,\ldots,Y_n$ is a trajectory from a $\mbox{B-GARCH}(1,1)$ initialised at $(Z_0,h_0)\sim\mu$, with $\mu$ being the stationary distribution. Then,\\
	(i) with probability tending to 1 as $n\rightarrow\infty$, there exists an open neighborhood set $V\subset\Theta$ of $\boldsymbol\theta$ such that the conditional log-likelihood function $\ell(\boldsymbol\theta)$ is concave in $V$, and then there is a unique maximum point $\widehat{\boldsymbol\theta}$ of $\ell(\boldsymbol\theta)$ in $V$ that solves ${\bf U}(\widehat{\boldsymbol\theta})=0$;\\
	(ii) $\widehat{\boldsymbol\theta}\stackrel{p}{\longrightarrow}\boldsymbol\theta$ as $n\rightarrow\infty$;\\
	(iii) $\widehat{\boldsymbol{\theta}}$ satisfies the asymptotic normality $\sqrt{n}(\widehat{\boldsymbol{\theta}}-\boldsymbol\theta)\stackrel{d}{\longrightarrow}N({\bf 0}, \boldsymbol\Sigma^{-1})$ as $n\rightarrow\infty$, where 
	\begin{eqnarray}\label{eq:Sigma}
	\boldsymbol\Sigma=\dfrac{1}{4}E\left[\left(\psi'\left(\frac{1/h_t-1}{2}\right)-\psi'\left(\frac{1}{2h_t}\right)\right)\dfrac{1}{h_t^4}\dfrac{\partial h_t}{\partial\boldsymbol\theta}\dfrac{\partial h_t}{\partial\boldsymbol\theta}^\top\right],
	\end{eqnarray}
with $\dfrac{\partial h_t}{\partial\boldsymbol\theta}=\left(\dfrac{\partial h_t}{\partial\alpha_0},\dfrac{\partial h_t}{\partial\alpha_1},\dfrac{\partial h_t}{\partial\beta_1}\right)^\top$.
\end{theorem}

To perform likelihood-based inference, we need to set initial values for the unobservable initial conditional variance to compute the conditional log-likelihood function. For instance, under the B-GARCH(1,1) model, we need to specify some initial value for $h_1$, say $\widetilde h_1$. Therefore, in practice, we replace $\ell(\boldsymbol{\theta})$ with the conditional log-likelihood function corresponding to this initialisation, which we denote by $\widetilde{\ell}(\boldsymbol{\theta})$. In the next theorem, we show that $\ell(\boldsymbol\theta)$ is well approximated by $\widetilde\ell(\boldsymbol\theta)$.

\begin{theorem}\label{thm:apporx_ell}
	Under the conditions of Theorem \ref{thm:AN},  
	$$\sup_{\boldsymbol\theta\in V}\left|\dfrac{1}{n}\ell(\boldsymbol\theta)-\dfrac{1}{n}\widetilde\ell(\boldsymbol\theta)\right|=o_p(1),$$
	with the set $V$ as defined there.
\end{theorem}

\section{Monte Carlo experiments}\label{sec:MC}

In this section, we conduct Monte Carlo (MC) simulations to empirically evaluate the performance of the conditional maximum likelihood estimators. We further assess their associated uncertainty quantification using the conditional information matrix defined in (\ref{information}). Trajectories are generated from a B-GARCH(1,1) model with conditional variance $h_t = \alpha_0 + \alpha_1 Y^2_{t-1} + \beta_1 h_{t-1}$, where the parameters are set to $\alpha_0 = 7\times10^{-7}$, $\alpha_1 = 0.36$, and $\beta_1 = 0.54$. We consider sample sizes of $n \in \{350, 500, 700, 1000\}$ with $1000$ Monte Carlo replications for each scenario.

\begin{figure}[ht!]
	\centering
	\includegraphics[width = 1\linewidth]{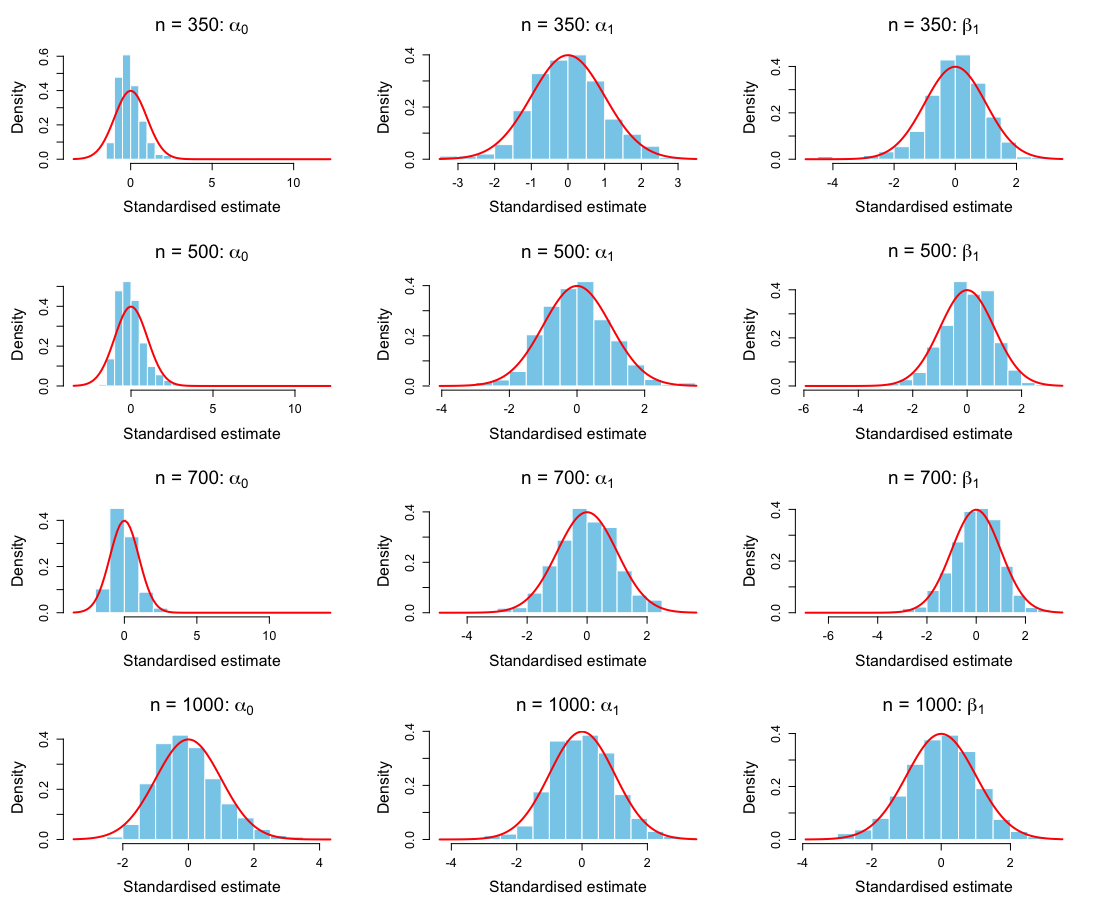}
	\caption{Histograms of the standardised (mean-standard deviation) Monte Carlo estimates along with the $N(0,1)$ density function. Trajectories generated from the B-GARCH(1,1) model with $\alpha_0 = 7\times10^{-7}$, $\alpha_1 = 0.36$, and $\beta_1 = 0.54$, and sample size $n \in \{350, 500, 700, 1000\}$.}\label{fig:histograms}
\end{figure}

Table \ref{tab:bgarch} summarises the Monte Carlo mean estimates alongside empirical and information-based standard errors for the B-GARCH(1,1) model. The results demonstrate that as the sample size $n$ increases, the estimators converge toward their true values, while the close alignment between empirical and information-based standard errors validates the reliability of the information matrix for asymptotic uncertainty quantification. This claim is also supported from the results presented in Table \ref{tab:coverage_results}, which provides the empirical coverage of asymptotic confidence intervals (CIs) for B-GARCH(1,1) parameters at $90\%$, $95\%$, and $99\%$ nominal levels. In other words, we compute the number of times that the CI include the true value of the parameter. The coverage results collectively demonstrate that the asymptotic properties of the information-based inference is increasingly satisfied as the sample size grows, with good empirical coverage rates for all parameters. Furthermore, Figure \ref{fig:histograms} displays the histograms of the MC estimates along with the $N(0,1)$ density function. We can observe a good Gaussian approximation for all cases, which improves as the sample size increases, and therefore is aligned with our theoretical results established in Theorem \ref{thm:AN} on the asymptotic normality. It is important to mention that extensive simulations were conducted under alternative parameter configurations. As these yielded qualitatively similar results, they are omitted here for brevity. These supplementary findings are available from the author upon request.

\begin{table}[ht!]
	\centering
	\setlength{\tabcolsep}{4pt} 
	\begin{tabular}{r ccc ccc ccc}
		\toprule
		& \multicolumn{3}{c}{$\alpha_0$ ($\times 10^{-7}$)} & \multicolumn{3}{c}{$\alpha_1$} & \multicolumn{3}{c}{$\beta_1$} \\
		\cmidrule(r){2-4} \cmidrule(lr){5-7} \cmidrule(l){8-10}
		$n$ & Mean & Emp SE & Info SE & Mean & Emp SE & Info SE & Mean & Emp SE & Info SE \\
		\midrule
		350  & 8.10 & 4.30 & 3.09 & 0.3552 & 0.0919 & 0.0825 & 0.5182 & 0.1059 & 0.1002 \\
		500  & 7.82 & 3.16 & 2.36 & 0.3576 & 0.0757 & 0.0699 & 0.5226 & 0.0880 & 0.0790 \\
		700  & 7.41 & 2.32 & 1.90 & 0.3558 & 0.0622 & 0.0590 & 0.5337 & 0.0698 & 0.0646 \\
		1000 & 7.40 & 1.63 & 1.57 & 0.3564 & 0.0521 & 0.0498 & 0.5340 & 0.0525 & 0.0535 \\
		\bottomrule
	\end{tabular}
	\caption{MC mean estimates and standard errors based on the MC (Emp SE) and on the information matrix (Info SE). Trajectories generated from the B-GARCH(1,1) model with $\alpha_0 = 7\times10^{-7}$, $\alpha_1 = 0.36$, and $\beta_1 = 0.54$, and sample size $n \in \{350, 500, 700, 1000\}$.}
\label{tab:bgarch}
\end{table}

\begin{table}[ht!]
	\centering
	\setlength{\tabcolsep}{3.5pt} 
	\begin{tabular}{@{}r ccc ccc ccc@{}}
		\toprule
		\multirow{2.5}{*}{$n$} & \multicolumn{3}{c}{$\alpha_0$} & \multicolumn{3}{c}{$\alpha_1$} & \multicolumn{3}{c}{$\beta_1$} \\
		\cmidrule(lr){2-4} \cmidrule(lr){5-7} \cmidrule(lr){8-10}
		& 90\% & 95\% & 99\% & 90\% & 95\% & 99\% & 90\% & 95\% & 99\% \\
		\midrule
		350  & 0.8970 & 0.9360 & 0.9760 & 0.8530 & 0.9110 & 0.9670 & 0.8760 & 0.9240 & 0.9770 \\
		500  & 0.8910 & 0.9440 & 0.9730 & 0.8700 & 0.9240 & 0.9670 & 0.8830 & 0.9330 & 0.9760 \\
		700  & 0.8810 & 0.9370 & 0.9760 & 0.8810 & 0.9250 & 0.9740 & 0.8800 & 0.9360 & 0.9750 \\
		1000 & 0.9110 & 0.9540 & 0.9920 & 0.8950 & 0.9430 & 0.9720 & 0.9080 & 0.9490 & 0.9870 \\
		\bottomrule
	\end{tabular}
	\caption{Empirical coverage of asymptotic confidence intervals for B-GARCH(1,1) parameters at $90\%$, $95\%$, and $99\%$ nominal levels. Trajectories generated from the B-GARCH(1,1) model with $\alpha_0 = 7\times10^{-7}$, $\alpha_1 = 0.36$, and $\beta_1 = 0.54$, and sample size $n \in \{350, 500, 700, 1000\}$.}
	\label{tab:coverage_results}
\end{table}

\section{Generalisation}\label{sec:generalisations}

We generalise the baseline B-GARCH framework to accommodate asymmetry by introducing a dynamic, time-varying conditional mean $\mu_t$ rather than assuming a zero-mean process. Specifically, we consider a beta distribution on the interval $(-1,1)$ parameterised by mean $\mu \in (-1,1)$ and variance $\sigma^2 \in (0,1)$, subject to the constraint $\mu^2 + \sigma^2 < 1$ to guarantee a valid distribution. We denote $Y \sim \text{Beta}(\mu, \sigma^2)$, with probability density function given by
\begin{equation}\label{asym_beta}
	f(y; \mu, \sigma^2) = \frac{(1+y)^{\kappa^+(\mu,\sigma^2) - 1} \, (1-y)^{\kappa^-(\mu,\sigma^2) - 1}}{2^{\frac{1 - \mu^2-\sigma^2}{\sigma^2}-1} \, B\left(\kappa^+(\mu,\sigma^2),\kappa^-(\mu,\sigma^2) \right)},\quad y\in(-1,1),
\end{equation}
where $\kappa^+(\mu,\sigma^2) = \left(\frac{1 + \mu}{2}\right)\left(\frac{1 - \mu^2-\sigma^2}{\sigma^2}\right)$, $\kappa^-(\mu,\sigma^2) = \left(\frac{1 - \mu}{2}\right)\left(\frac{1 - \mu^2 - \sigma^2}{\sigma^2}\right)$, and $B(\cdot,\cdot)$ denotes the beta function. Setting $\mu = 0$ in \eqref{asym_beta} recovers the symmetric beta density in \eqref{density}. Next, we introduce the $\operatorname{AR}(1)\operatorname{-B-GARCH}(1,1)$ specification, which jointly models conditional mean and variance dynamics.

\begin{definition}\label{def:modelext}[$\operatorname{AR}(1)\operatorname{-B-GARCH}(1,1)$ processes]
	We say that a stochastic sequence $\{Y_t\}_{t\in \mathbb{Z}}$ is an $\operatorname{AR}(1)\operatorname{-B-GARCH}(1,1)$ process if 
	\begin{equation*}
		Y_t \mid \mathcal{F}_{t-1} \sim \operatorname{Beta}(\mu_t,h_t), \quad \forall t \geq 1,
	\end{equation*}
	where $\mathcal{F}_{t-1}=\sigma(Y_{t-1}\mu_{t-1},h_{t-1},Y_{t-2},\mu_{t-2},h_{t-2},\ldots)$ is the filtration generated by the past history, and the conditional mean $\mu_t=E(Y_t\mid \mathcal{F}_{t-1})$  and conditional variance $h_t=\mbox{Var}(Y_t\mid \mathcal{F}_{t-1})$ evolve according to
			\begin{equation*}\label{ht:condmean_gen}
			\mu_t = \rho_0 + \rho_1 Y_{t-1}, \quad \forall t \geq 1,
		\end{equation*}	
	and
		\begin{equation}\label{ht:condvariance_gen}
		h_t = \alpha_0 +  \alpha_1 g(\epsilon_{t-1}) + \beta_1 h_{t-1}, \quad \forall t \geq 1,
	\end{equation}	
	respectively, where $\epsilon_{t-1}=Y_{t-1}-\mu_{t-1}$, and $g:(0,1)\rightarrow(0,1)$, and the parameters satisfy $|\rho_0|+|\rho_1|<1$, $\alpha_0>0$, $\alpha_1 \geq 0$, $\beta_1 \geq 0$, and the strict upper bound constraint 
	\begin{equation*}\label{const_gen}
		|\rho_0|+|\rho_1|+\alpha_0 +  \alpha_1 + \beta_1 < 1.
	\end{equation*}
\end{definition}

The symmetric B-GARCH(1,1) model is recovered as a special case of Definition~\ref{def:modelext} by setting $\rho_0 = \rho_1 = 0$ and specifying $g$ as the identity function. In our empirical application in the next section, we consider two specifications for $g(\epsilon_{t-1})$ in (\ref{ht:condvariance_gen}): the scaled quadratic function $g(\epsilon_{t-1}) = \epsilon_{t-1}^2/4$ and the exponential function $g(\epsilon_{t-1}) = 1 - \exp(-\gamma \epsilon_{t-1}^2)$ ($\gamma > 0$), with the latter specification being inspired by the continuous smooth-transition GARCH models by \cite{riv1998}.  The scaled quadratic specification normalises $\epsilon_{t-1}^2 \in [0, 4)$ onto $[0, 1)$, preserving classical quadratic ARCH dynamics while guaranteeing $h_t \in (0, 1)$ almost surely. The exponential specification similarly bounds $g(\cdot) \in [0, 1)$ but introduces non-linear saturation: it acts quadratically for small shocks ($g(\epsilon_{t-1}) \approx \gamma \epsilon_{t-1}^2$) and approaches $1$ asymptotically for extreme innovations, preventing large boundary shocks from over-inflating conditional variance.
	
\begin{remark}
	\label{rem:ar_b_garch_extensions}
	Strict stationarity and ergodicity for the generalised $\text{AR-B-GARCH}$ processes can be established by adapting the proof strategy of Theorem~\ref{thm:staterg}, which relies on verifying the contraction conditions of \citet{douwin2008}. Parameter estimation proceeds via conditional maximum likelihood estimation, with asymptotic consistency and normality following analogously to the baseline $\text{B-GARCH}$ specification in Theorem~\ref{thm:AN}.
\end{remark}

\section{Realised correlation application}\label{sec:app}

We apply our bounded GARCH processes for modelling monthly realised correlation. We collected the data from \url{finance.yahoo.com}, which consists of historical daily adjusted closing prices for the Vanguard Real Estate ETF (VNQ) and the iShares 20+ Year Treasury Bond ETF (TLT) over the sixteen-year period from January 1, 2010, to December 31, 2025. Daily log-returns are constructed for both assets to calculate a non-overlapping monthly realised correlation time series, say $y_t \in (-1, 1)$, defined as the intra-month sample Pearson correlation coefficient across daily returns for every calendar month. The resulting time series comprises $n = 192$ monthly observations with a sample mean of $0.0210$ and historical values ranging from a minimum of $-0.8038$ to a maximum of $0.7719$. This pair provides an ideal empirical benchmark for bounded models because it exhibits wide dynamic swings across both positive and negative regimes while repeatedly approaching the theoretical boundaries of $(-1, 1)$. Under such severe regime shifts and boundary proximity, standard unbounded models, such as conventional Gaussian GARCH specifications, suffer severe structural deficiencies by generating impossible predictive intervals that breach domain limits (exceeding $+1.0$ or falling below $-1.0$) and failing to capture the natural conditional variance contraction that occurs near extreme boundaries, whereas our bounded model specifications dynamically adjust their predictive regions to preserve domain feasibility without sacrificing empirical coverage, as it will be illustrated in the sequence. 

\begin{figure}[ht!]
	\centering
	\includegraphics[width = 1\linewidth]{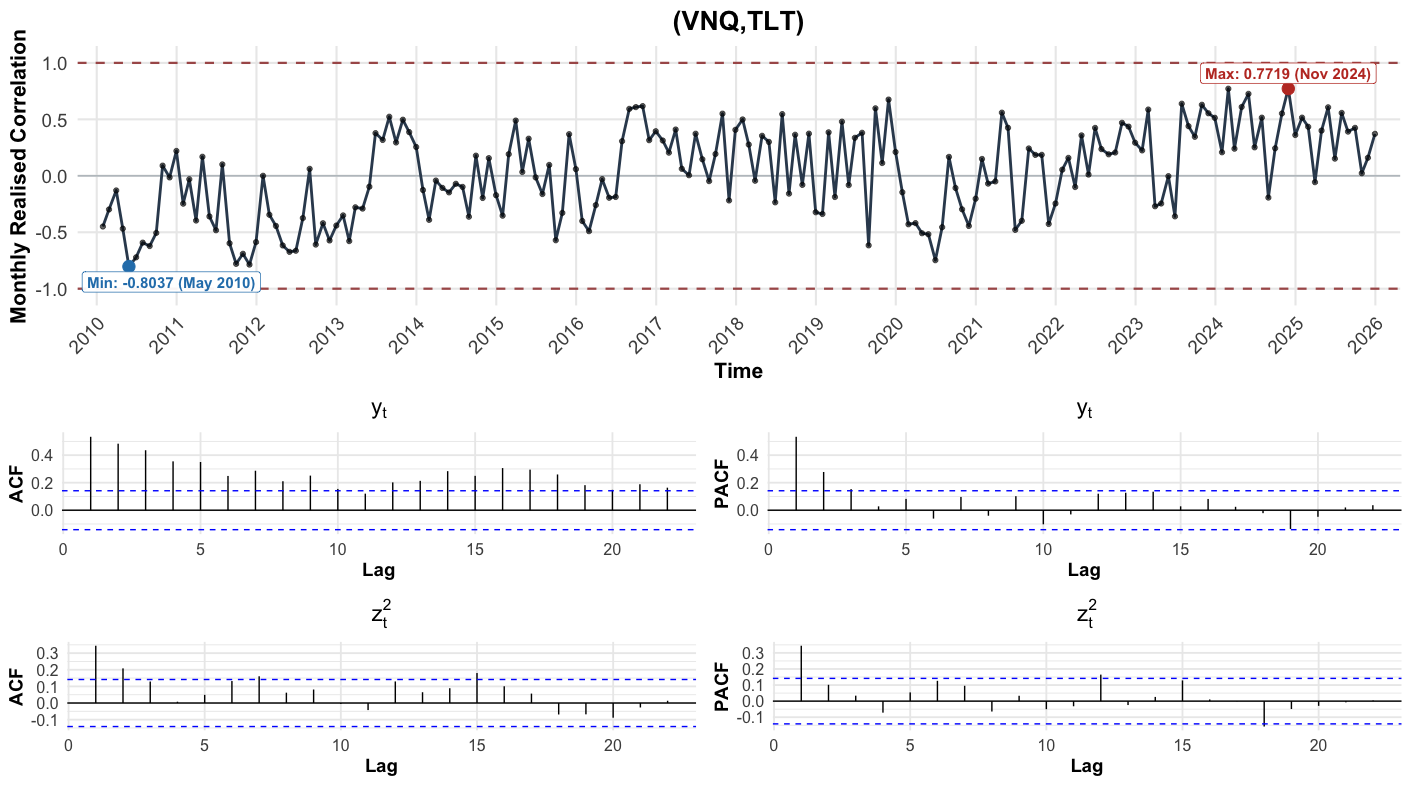}
	\caption{Monthly realised correlation $y_t$ for (VNQ, TLT) ($n = 192$, Jan 2010 - Dec 2025) with minimum and maximum callouts (top), sample ACF/PACF of $y_t$ (middle), and sample ACF/PACF of squared demeaned deviations $z_t = (y_t - \bar{y})^2$ (bottom).}\label{fig:RC_VNQxTLT_plot}
\end{figure}

Figure~\ref{fig:RC_VNQxTLT_plot} illustrates the monthly realised correlation time series $y_t$, its sample autocorrelation (ACF) and partial autocorrelation (PACF) functions, and the corresponding sample autocorrelations for the squared demeaned series $z_t = (y_t - \bar{y})^2$, where $\bar{y} = n^{-1}\sum_{t=1}^{192} y_t$. The series exhibits pronounced time dependence in both its conditional mean and conditional variance (GARCH-type volatility clustering), confirming the necessity of joint autoregressive and conditional variance modelling.

We evaluate three AR(1)-B-GARCH specifications applied to the monthly realised correlation time series: the AR(1)-B-GARCH(1,0) model with a quadratic shock $g(\epsilon_{t-1}) = \epsilon_{t-1}^2 / 4$, the AR(1)-B-GARCH(0,1) model ($\alpha_1 = 0$), and the AR(1)-B-GARCH(1,0) model with an exponential shock $g(\epsilon_{t-1}) = 1 - \exp(-\gamma \epsilon_{t-1}^2)$. Table~\ref{tab:summaryfit} reports the conditional maximum likelihood estimates, along with standard errors and $95\%$ percentile confidence intervals calculated from $B = 500$ parametric bootstrap replications. For comparison, we fit a standard Gaussian AR(1)-GARCH(1,1) model given by
\begin{align}
	y_t &= \underset{(0.0241)}{0.0156} + \underset{(0.0643)}{0.5597} \, y_{t-1} + \varepsilon_t, \label{eq:bench_mean} \\
	\varepsilon_t &= \sigma_t \eta_t, \quad \eta_t \overset{\text{i.i.d.}}{\sim} N(0, 1), \notag \\
	\sigma_t^2 &= \underset{(0.0145)}{0.0135} + \underset{(0.0626)}{0.0743} \, \varepsilon_{t-1}^2 + \underset{(0.1620)}{0.8024} \, \sigma_{t-1}^2, \label{eq:bench_var}
\end{align}
where the values in parentheses denote standard errors.

\begin{table}[htbp]
	\centering
	\begin{tabular}{llccc}
		\toprule
		Model & Parameter & Estimate & Std. Error & 95\% Percentile CI \\
		\midrule
		\multirow{4}{*}{\begin{tabular}{@{}l@{}}1. AR(1)-B-GARCH(1,0) \\ $g(\epsilon_{t-1})=\epsilon_{t-1}^2/4$\end{tabular}}
		& $\rho_0$   & \phantom{-}0.0117 &    0.0237 & $[-0.0306,\ 0.0596]$ \\
		& $\rho_1$   & \phantom{-}0.5121 &    0.0630 & $[\phantom{-}0.3567,\ 0.6108]$ \\
		& $\alpha_0$ & \phantom{-}0.1022 &    0.0091 & $[\phantom{-}0.0820,\ 0.1171]$ \\
		& $\alpha_1$ & \phantom{-}0.1066 &    0.1530 & $[\phantom{-}0.0005,\ 0.4476]$ \\
		\midrule
		\multirow{4}{*}{2. AR(1)-B-GARCH(0,1)} 
		& $\rho_0$   & 0.0000          &    0.0215 & $[-0.0445,\ 0.0427]$ \\
		& $\rho_1$   & \phantom{-}0.5087 &    0.0616 & $[\phantom{-}0.3653,\ 0.6051]$ \\
		& $\alpha_0$ & \phantom{-}0.0604 &    0.0104 & $[\phantom{-}0.0486,\ 0.0865]$ \\
		& $\beta_1$  & \phantom{-}0.4270 &    0.0087 & $[\phantom{-}0.2100,\ 0.5152]$ \\
		\midrule
		\multirow{5}{*}{\begin{tabular}{@{}l@{}}3. AR(1)-B-GARCH(1,0) \\ $g(\epsilon_{t-1})=1-\exp\left(-\gamma\epsilon_{t-1}^2\right)$\end{tabular}}
		& $\rho_0$   & \phantom{-}0.0115 &    0.0246 & $[-0.0316,\ 0.0618]$ \\
		& $\rho_1$   & \phantom{-}0.5120 &    0.0614 & $[\phantom{-}0.3723,\ 0.6010]$ \\
		& $\alpha_0$ & \phantom{-}0.1025 &    0.0143 & $[\phantom{-}0.0718,\ 0.1201]$ \\
		& $\alpha_1$ & \phantom{-}0.0879 &    0.1058 & $[\phantom{-}0.0012,\ 0.3565]$ \\
		\bottomrule
	\end{tabular}
\caption{Parameter estimates obtained via conditional maximum likelihood estimation, and standard errors and 95\% percentile confidence intervals obtained via parametric bootstrap ($B=500$ replications) for three AR-B-GARCH models applied to the realised correlation. The estimate of the $\gamma$ parameter for Model 3 is $\widehat\gamma=0.2893 $.}\label{tab:summaryfit}
\end{table}

The autoregressive parameter estimates for the conditional mean $\rho_1$ are remarkably consistent across all three AR-B-GARCH specifications, taking values near $0.51$ ($\widehat{\rho}_1 = 0.5121$, $0.5087$, and $0.5120$ for Models 1--3, respectively). The intercept $\rho_0$ remains close to zero, with $95\%$ bootstrap confidence intervals containing zero in every specification. Concerning conditional variance dynamics, Model 1 yields ARCH estimates of $\widehat{\alpha}_0 = 0.1022$ and $\widehat{\alpha}_1 = 0.1066$. In Model 2 ($\alpha_1 = 0$), the variance persistence parameter is estimated as $\widehat{\beta}_1 = 0.4270$ with a bootstrap interval of $[0.2100, 0.5152]$. For Model 3, joint estimation of the saturation rate yields $\widehat{\gamma} = 0.2893$ alongside $\widehat{\alpha}_1 = 0.0879$, where the exponential shock transformation $g(\epsilon_{t-1}) = 1 - \exp(-\gamma \epsilon_{t-1}^2) \in [0, 1)$ strictly bounds the impact of large innovations on the conditional variance $h_t$. Parameter uncertainty is quantified via a parametric bootstrap scheme ($B = 500$) because non-negativity constraints near the boundary ($\alpha_1 \ge 0, \beta_1 \ge 0$) might violate standard asymptotic normality approximations under small sample sizes.

Although the benchmark Gaussian AR(1)-GARCH(1,1) model in Equations~\eqref{eq:bench_mean}--\eqref{eq:bench_var} exhibits similar mean persistence ($\widehat{\rho}_1 = 0.5597$), its conditional variance structure is unconstrained with respect to the support of $y_t$. Because the Gaussian distribution assigns positive probability mass outside $(-1, 1)$, its predictive interval bounds systematically exceed theoretical correlation limits during volatile periods. In contrast, the bounded GARCH framework ensures strict domain compliance by modelling $y_t$ directly on $(-1, 1)$, with the conditional variance naturally bounded above by $1 - \mu_t^2$ ($\mu_t^2+h_t<1$).

\begin{figure}[ht!]
	\centering
	\includegraphics[width = 1\linewidth]{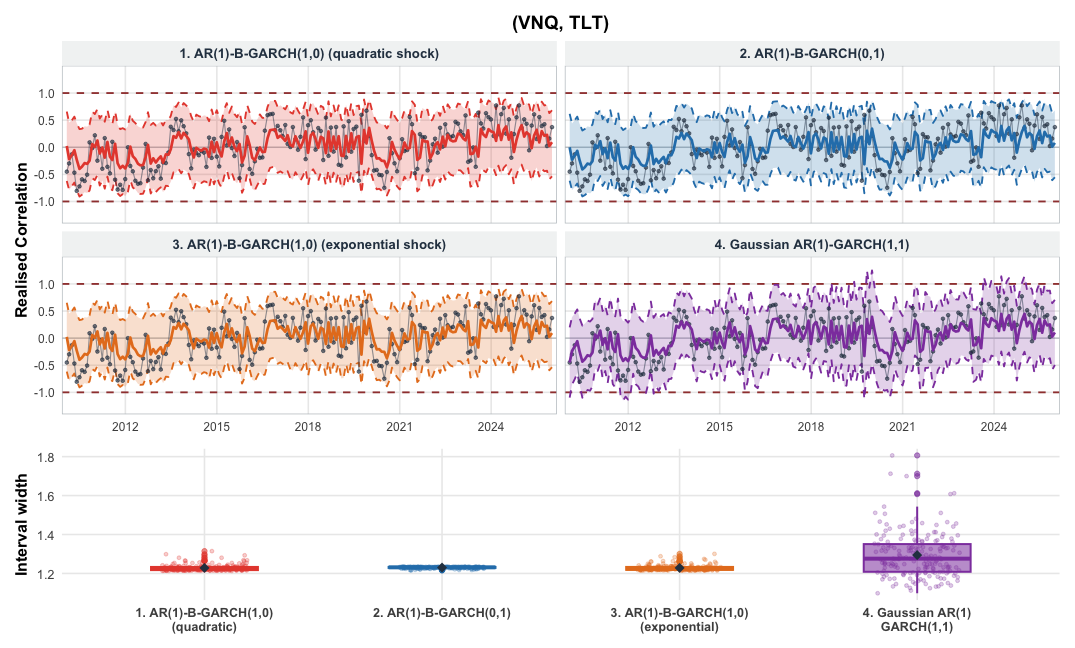}
\caption{In-sample $95\%$ predictive regions and conditional mean estimates for the monthly realised correlation between VNQ and TLT across four model specifications: AR(1)-B-GARCH(1,0) with quadratic shock, AR(1)-B-GARCH(0,1) ($\alpha_1 = 0$), AR(1)-B-GARCH(1,0) with exponential shock, and the benchmark Gaussian AR(1)-GARCH(1,1). Top panels: points depict observed realized correlations $y_t$, solid lines denote fitted conditional means, and shaded regions represent $95\%$ predictive intervals. Dashed red lines mark theoretical domain bounds at $y = \pm 1$. Bottom panel: boxplots displaying the empirical distribution of $95\%$ predictive interval widths across models, with diamonds representing sample means.}\label{fig:VNQxTLT_pred_plot}
\end{figure}

Table~\ref{tab:empirical_perf} summarises the Akaike Information Criterion (AIC), empirical coverage rates of in-sample $95\%$ confidence intervals, and domain breach percentages for all models considered. In terms of relative fit, all three bounded GARCH specifications demonstrate clear superiority over the Gaussian benchmark, yielding lower AIC values. While empirical coverage is well-calibrated near the nominal $95\%$ level across all models ($94.27\%$--$94.79\%$), domain support differs fundamentally. The Gaussian specification suffers from structural misspecification, incurring a $12.50\%$ domain breach rate where predictive intervals spill outside $(-1, 1)$. Conversely, all AR-B-GARCH processes exhibit no domain violations by construction. As illustrated in Figure~\ref{fig:VNQxTLT_pred_plot}, the classic Gaussian AR(1)-GARCH(1,1) model yields excessively wide predictive intervals during high-volatility regimes, whereas the proposed bounded specifications adaptively contract near the boundaries $(-1, 1)$ to preserve theoretical domain integrity.

\begin{table}[ht!]
	\centering
	\begin{tabular}{lccc}
		\toprule
	Model Specification & AIC & 95\% Coverage & Boundary Violations \\
		\midrule
		1. AR(1)-B-GARCH(1,0) (quadratic shock) & $111.62$ & $94.27\%$ & $0.00\%$ \\
		2. AR(1)-B-GARCH(0,1) ($\alpha_1 = 0, \beta_1 > 0$) & $111.96$ & $94.79\%$ & $0.00\%$ \\
		3. AR(1)-B-GARCH(1,0) (exponential shock) & $113.64$ & $94.27\%$ & $0.00\%$ \\
		\midrule
		4. Gaussian AR(1)-GARCH(1,1) & $126.82$ & $94.79\%$ & $12.50\%$ \\
		\bottomrule
	\end{tabular}
			\caption{Empirical performance comparison across bounded AR-B-GARCH models and the Gaussian benchmark.}
\label{tab:empirical_perf}
\end{table}

\begin{figure}[ht!]
	\centering
	\includegraphics[width = 1\linewidth]{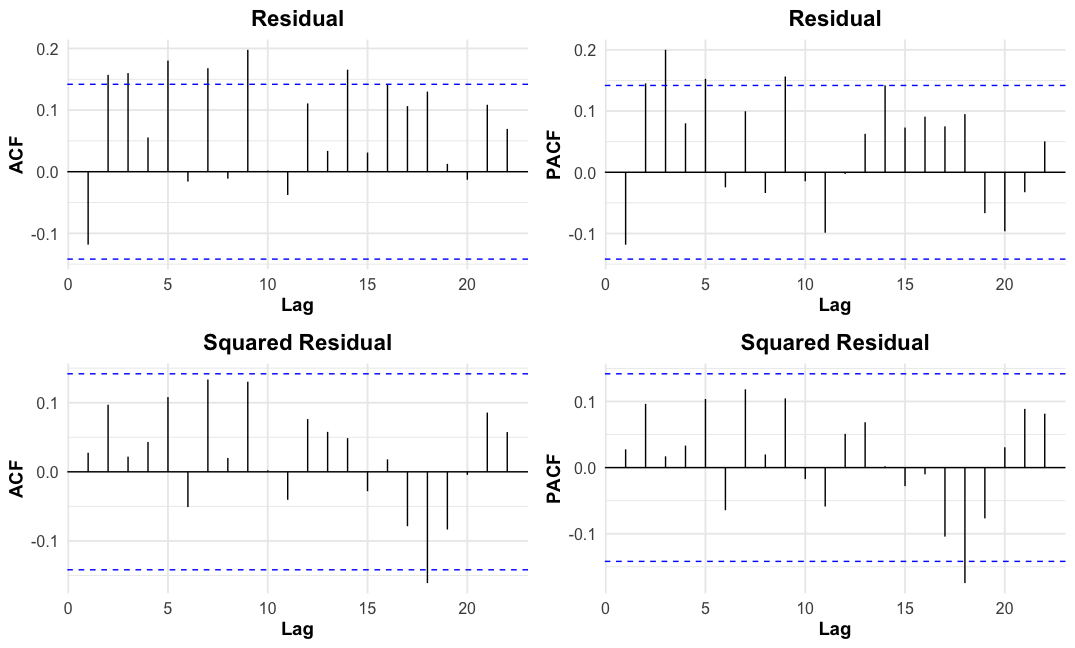}
	\caption{ACF and PACF of the Pearson residuals for the AR(1)-B-GARCH(1,0) (quadratic shock) model.}\label{fig:res_analysis}
\end{figure}

Finally, Figure~\ref{fig:res_analysis} presents the sample autocorrelation (ACF) and partial autocorrelation (PACF) functions of the standardised Pearson residuals and squared Pearson residuals for the AR(1)-B-GARCH(1,0) model with quadratic shock. Diagnostic plots for the remaining bounded GARCH processes exhibit virtually identical features and are omitted for brevity. The absence of significant residual autocorrelation confirms that the proposed bounded framework adequately captures the temporal dependence in both the conditional mean and conditional variance processes.

\section{Conclusion and future research}\label{sec:conclusion}

This paper introduced the Bounded GARCH (B-GARCH) class of processes, specifically designed to model the conditional heteroskedasticity of continuous time series restricted to a compact interval. By utilising a reparameterised beta distribution on the support $(-1, 1)$, our framework directly models the dynamics of the conditional variance while strictly preserving the bounded nature of the process. This approach fundamentally resolves the support mismatch and subsequent boundary violations that plague classical GARCH specifications when applied to bounded data. Furthermore, by operating directly on the natural support of the data, the B-GARCH model bypasses the need for non-linear link-function transformations, which are commonly used in practice but distort interpretability and introduce estimation biases for quantities of interest. Theoretically, we established the existence of a unique strictly stationary solution and proved the ergodicity of the B-GARCH(1,1) process. We also addressed parameter estimation via conditional maximum likelihood and rigorously derived the associated consistency and asymptotic normality of the estimators. Our Monte Carlo simulation results validated these asymptotic findings, demonstrating that the estimators converge toward their true values as the sample size increases. We also proposed a generalisation of the B-GARCH processes to accommodate asymmetry, by incorporating a dynamic conditional mean rather than a zero-mean process specification. We presented an empirical illustration on realised asset correlations between Vanguard Real Estate ETF (VNQ) and the iShares 20+ Year Treasury Bond ETF (TLT) from January 1, 2010, to December 31, 2025 that demonstrated the superior predictive performance of bounded GARCH models while highlighting the boundary violations produced by the classical GARCH specification.

\begin{figure}[ht!]
	\centering
	\includegraphics[width = 1\linewidth]{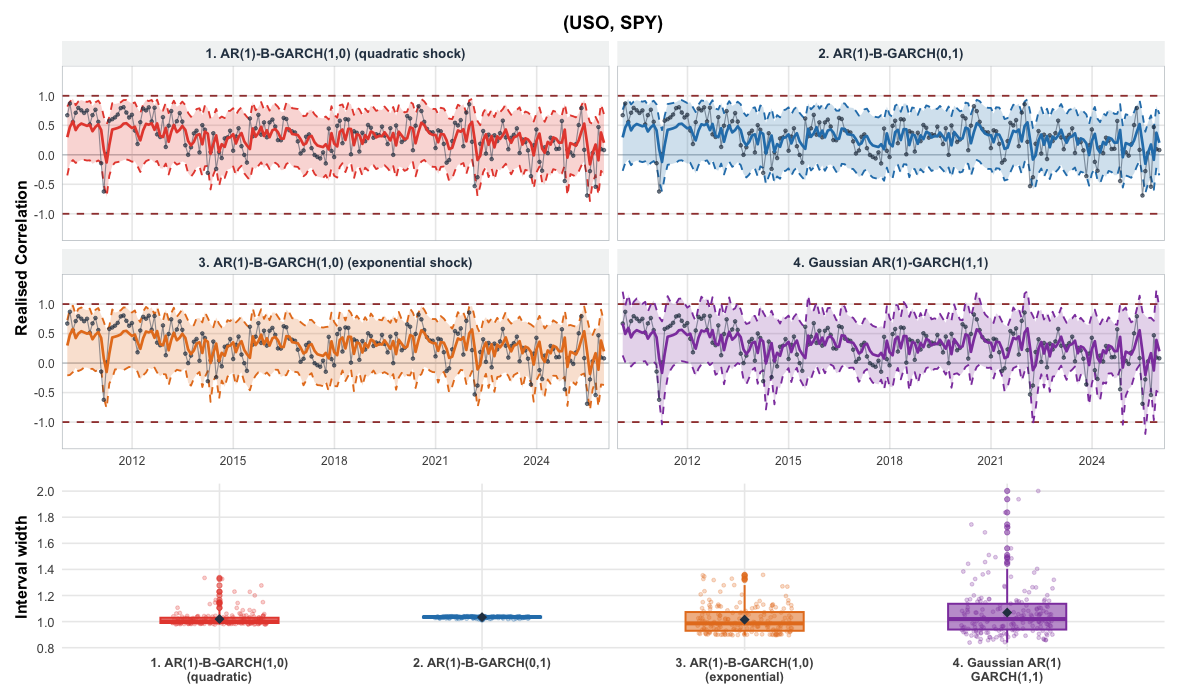}
		\includegraphics[width = 1\linewidth]{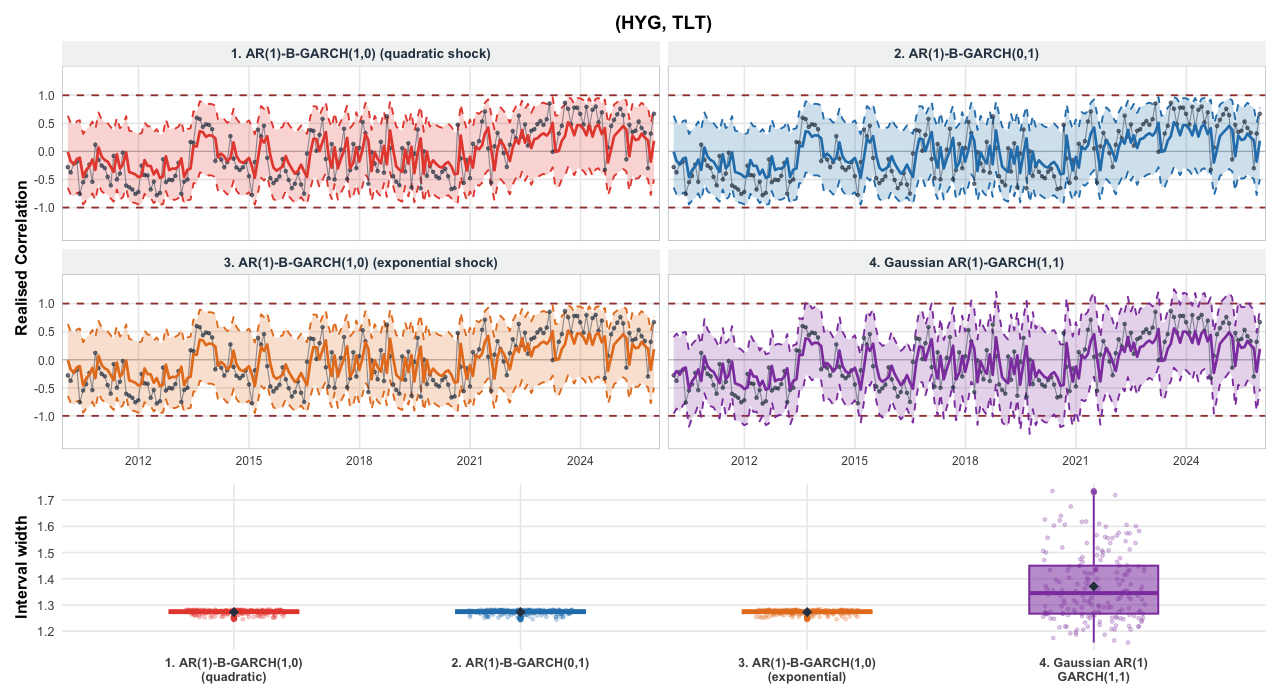}
	\caption{In-sample $95\%$ predictive regions and conditional mean estimates for the monthly realised correlation between USO and SPY and between HYG and TLT across four model specifications.}\label{fig:others_pred_plot}
\end{figure}

To assess robustness across broader empirical settings, we evaluated additional asset correlation pairs. As shown in Figure \ref{fig:others_pred_plot}, we construct $95\%$ in-sample predictive intervals and conditional mean estimates for the monthly realised correlation between the United States Oil Fund (USO) and the SPDR S\&P 500 ETF (SPY), as well as between the iShares High Yield Corporate Bond ETF (HYG) and the iShares 20+ Year Treasury Bond ETF (TLT), across the four candidate specifications. While the classical Gaussian AR-GARCH model yields severe boundary violations, exceeding theoretical bounds in $19.27\%$ and $38.02\%$ of observations for the (USO, SPY) and (HYG, TLT) pairs, respectively, the proposed B-GARCH models maintain support preservation across the entire sample while accurately capturing dynamic heteroskedasticity. Although our empirical application focuses on financial correlations, the proposed framework is broadly applicable to any bounded time series. Promising domains of application span macroeconomics and labor economics (unemployment rates), public finance (budget expenditure shares), financial econometrics (bounded market indices), political science (political approval ratings), and climatology (relative humidity).

Several avenues remain for future research. Investigating the kurtosis properties of the B-GARCH(1,0) process under the condition $\alpha_0 < \frac{1-\alpha_1}{2}$ would provide deeper insights into the tail behavior of the model. Applying the B-GARCH framework to other bounded time series, such as expenditure fractions and bounded financial indices, presents a natural next step to systematically evaluate the model's empirical scalability. We finish this section by proposing an extension of the B-GARCH models.

In Definition \ref{def:modeldel}, specifically in Equation (\ref{ht:condvariance}), the parameters are constrained to be positive and their sum must be less than 1 to ensure that the conditional variance of the B-GARCH model is well-posed. These restrictions can be relaxed by introducing a monotonic continuous link function $v:(0,1)\rightarrow \mathbb{R}$ into the conditional variance formulation, as follows:
\begin{equation}\label{ht:condvariance_extension}
	v(h_t) = \alpha_0 + \sum_{i=1}^p \alpha_i v(Y^2_{t-i}) + \sum_{j=1}^q \beta_j v(h_{t-j}), \quad \forall t \geq 1,
\end{equation}	
where the parameters $\alpha_0$, $\alpha_i $ for $i=1,\ldots,p$, and $\beta_j $ for $j=1,\ldots,q$ are real-valued. Common choices for $v$ include the logit and probit functions, yielding what we refer to as the logit or probit B-GARCH model, respectively. The standard model is recovered from this general formulation by applying the identity link $v(x)=x$ for $x\in(0,1)$; however, because this function does not map into $\mathbb{R}$, the parameter constraints must be reapplied. Ultimately, Model (\ref{ht:condvariance_extension}) provides greater flexibility; for instance, by allowing for negative autocorrelation, which is unattainable under the standard formulation.

Furthermore, this generalisation easily facilitates the incorporation of exogenous covariates. In this case, the conditional variance takes the form:
\begin{equation*}
	v(h_t) = \alpha_0 + \sum_{i=1}^p \alpha_i v(Y^2_{t-i}) + \sum_{j=1}^q \beta_j v(h_{t-j}) + {\bf Z}_t^\top\boldsymbol\gamma, \quad \forall t \geq 1,
\end{equation*}	
where ${\bf Z}_t$ is a sequence of covariate vectors, $\boldsymbol\gamma$ is the associated vector of regression coefficients, and the remaining quantities are defined as above.\\

\section*{Appendix}

\begin{proof}{\it [Proposition \ref{prop:kurtosis}]}.\\
	(i) It follows straightforwardly by using the moments of the $\mbox{Beta}(0,\sigma^2)$ distribution provided after Eq. (\ref{density}). \\
	(ii) We have that $E(Y_t^4|\mathcal F_{t-1})=\dfrac{3h_t^2}{2h_t+1}$. By using Jensen's inequality, it follows that 
	\begin{eqnarray}\label{proof1.1:ineq}
		E(Y_t^4)\geq\dfrac{3E(h_t^2)}{2E(h_t)+1}=\dfrac{\alpha_0^2+\frac{2\alpha_1\alpha_0^2}{1-\alpha_1}+\alpha_1^2E(Y_t^4)}{\frac{2\alpha_0}{1-\alpha_1}},
	\end{eqnarray}
	where we have used that $E(h_t)=E(Y^2_{t-1})=\mbox{Var}(Y_t)=\dfrac{\alpha_0}{1-\alpha_1}$ and $E(Y_{t-1}^4)=E(Y_t^4)$ due to stationarity. We now rearrange the terms in inequality (\ref{proof1.1:ineq}) to obtain that 
	$$E(Y_t^4)\geq\dfrac{3\alpha_0^2(1+\alpha_1)}{2\alpha_0+(1-\alpha_1)(1-3\alpha_1^2)}.$$
	Hence,
	\begin{eqnarray*}
		\mbox{Kurt}(Y_t)=\dfrac{E(Y_t^4)}{[\mbox{Var}(Y_t)]^2}\geq \dfrac{(1-\alpha_1^2)}{\alpha_0^2}\dfrac{3\alpha_0^2(1+\alpha_1)}{2\alpha_0+(1-\alpha_1)(1-3\alpha_1^2)}=
		\dfrac{3(1-\alpha_1)(1-\alpha_1^2)}{2\alpha_0+(1-\alpha_1)(1-3\alpha_1^2)}.
	\end{eqnarray*}
	(iii) The forth conditional moment can be expressed as $E(Y_t^4|\mathcal F_{t-1})=\dfrac{3}{4}\left(2h_t-1+\dfrac{1}{2h_t+1}\right)$. Now, we use that $\dfrac{1}{2h_t+1}\leq\dfrac{1}{2\alpha_0+1}$ to obtain 
	$$E(Y_t^4)\leq\dfrac{3}{4}\left(\dfrac{2\alpha_0}{1-\alpha_1}-1+\dfrac{1}{2\alpha_0+1}\right)=\kappa_{noise}\dfrac{\alpha_0^2+\alpha_0\alpha_1/2}{1-\alpha_1},$$
	where $\kappa_{noise}$ is the kurtosis of the beta noise process. Hence, it follows that
	\begin{eqnarray*}
		\mbox{Kurt}(Y_t)&=&\dfrac{E(Y_t^4)}{[\mbox{Var}(Y_t)]^2}\leq\dfrac{(1-\alpha_1)^2}{\alpha_0^2}\kappa_{noise}\dfrac{\alpha_0^2+\alpha_0\alpha_1/2}{1-\alpha_1}\\
		&=&\kappa_{noise}(1-\alpha_1)\left(1+\dfrac{\alpha_1}{2\alpha_0}\right)\leq\kappa_{noise},
	\end{eqnarray*}	
	where the last inequality follows from the assumption $\alpha_0\geq\dfrac{1-\alpha_1}{2}$.
\end{proof}

\begin{proof}{\it [Theorem \ref{thm:mds}]}.  We start by computing the mean of $\log(1-Y^2)$, where $Y\sim\mbox{Beta}(0,h)$ (since the conditional distribution of $Y_t|\mathcal F_{t-1}$ is $\mbox{Beta}(0,h_t)$), which can be obtained via cumulant generating function. For $s\in\mathbb R$, it follows that
	\begin{eqnarray*}
		E\left(\exp\{s\log(1-Y^2)\}\right)&=&E\left((1-Y^2)^s\right)=\dfrac{\Gamma\left(\frac{1}{2h}\right)}{\sqrt{\pi}\Gamma\left(\frac{1/h-1}{2}\right)}\int_{-1}^1(1-y^2)^{(1/h+2s-3)/2}dy\\
		&=&\dfrac{\Gamma\left(\frac{1}{2h}\right)}{\Gamma\left(\frac{1/h-1}{2}\right)}
		\dfrac{\Gamma\left(\frac{1/h-1}{2}+s\right)}{\Gamma\left(\frac{1}{2h}+s\right)}.
	\end{eqnarray*}	
	Let $C(s)\equiv\log E\left(\exp\{s\log(1-Y^2)\}\right)$, that is, the cumulant generating function of $\log(1-Y^2)$. Then,
	\begin{eqnarray}\label{expr_condmean}
		E\left(\log(1-Y^2)\right)=\dfrac{dC(s)}{ds}\bigg|_{s=0}=\psi\left(\dfrac{1/h-1}{2}\right)-\psi\left(\frac{1}{2h}\right).	
	\end{eqnarray}	
	
	Using (\ref{expr_condmean}), we obtain that $E(U_{t,k}|\mathcal F_{t-1})=0$ for all $k$, with $U_{t,k}$ as defined in Equation (\ref{eq:Utk}). Let us now show that 
	$U_{t,k}$ is integrable to prove the MDS claim. It follows that
	\begin{eqnarray}\label{ineq:condexp}
		E(|U_{t,k}|)=E[E(U_{t,k}|\mathcal F_{t-1})]\leq E\left[\left(\psi\left(\frac{1}{2h_t}\right)-\psi\left(\dfrac{1/h_t-1}{2}\right)\right)\dfrac{1}{h^2_t}
		\left|\dfrac{\partial h_t}{\partial\theta_k}\right|\right],
	\end{eqnarray}	
	where we have used that $$E\left(\big|-\log(1-Y_t^2)-\psi\left(\frac{1}{2h_t}\right)+\psi\left(\frac{1/h_t-1}{2}\right)\big|\bigg|\mathcal F_{t-1}\right)\leq2\left(\psi\left(\frac{1}{2h_t}\right)-\psi\left(\frac{1/h_t-1}{2}\right)\right),$$ which is positive since $\psi(\cdot)$ is an increasing function, 
	$$\dfrac{\partial h_t}{\partial\alpha_0}=\dfrac{1-\beta_1^{t-1}}{1-\beta_1},\quad \dfrac{\partial h_t}{\partial\alpha_1}=\sum_{j=1}^{t-1}\beta_1^{j-1}Y^2_{t-j},$$
	and
	$$\dfrac{\partial h_t}{\partial\beta_1}=\alpha_0\dfrac{1-\beta_1^{t-1}-(t-1)\beta_1^{t-2}(1-\beta_1)}{(1-\beta_1)^2}+\alpha_1\sum_{j=2}^{t-1}(j-1)\beta_1^{j-2}Y^2_{t-j}+(t-1)\beta_1^{t-2}h_1.$$
	
	From the model definition, we have that $Y_t^2\in(0,1)$, $\alpha_0\in(0,1)$, $\alpha_1,\beta_1\in[0,1)$, and $\alpha_0\leq h_t\leq \alpha_0+\alpha_1+\beta_1$, with $\alpha_0+\alpha_1+\beta_1<1$. Hence, it can be checked that the quantities $\psi\left(\frac{1}{2h_t}\right)-\psi\left(\frac{1/h_t-1}{2}\right)$,  $1/h^2_t$, and $\left|\frac{\partial h_t}{\partial\theta_k}\right|$, involved on the right side of (\ref{ineq:condexp}) are bounded above, and therefore $E(|U_{t,k}|)<\infty$ for all $k$ and $t$.
	
	We now argue the square-integrability of ${\bf U}_t(\boldsymbol\theta)$. By using the cumulant generating function provided above, we obtain that that the variance of $\log(1-Y^2)$ with $Y\sim\mbox{Beta}(0,h)$ is 
	$$\mbox{Var}(\log(1-Y^2))=\dfrac{d^2C(s)}{ds^2}\bigg|_{s=0}=\psi'\left(\frac{1/h-1}{2}\right)-\psi'\left(\frac{1}{2h}\right).$$
	
	The above expression will be used to compute $\mbox{Var}(\log(1-Y^2_t)|\mathcal F_{t-1})$ in what follows. The second moment of $U_{t,k}$ can be expressed by
	\begin{eqnarray}\label{eq:condvarU}
		E(U_{t,k}^2)=\mbox{Var}(U_{t,k})=E(\mbox{Var}(U_{t,k})|\mathcal F_{t-1})+\mbox{Var}(E(U_{t,k})|\mathcal F_{t-1})=E(\mbox{Var}(U_{t,k})|\mathcal F_{t-1})+0=\nonumber\\
		E\left[\mbox{Var}(\log(1-Y_t^2)|\mathcal F_{t-1})\dfrac{1}{4h_t^4}\left(\dfrac{\partial h_t}{\partial\theta_k}\right)^2\right]	
		=E\left[\left(\psi'\left(\frac{1/h_t-1}{2}\right)-\psi'\left(\frac{1}{2h_t}\right)\right)\dfrac{1}{4h_t^4}\left(\dfrac{\partial h_t}{\partial\theta_k}\right)^2\right]. 
	\end{eqnarray}	
	
	Arguing similarly as above, the quantities in the expectation on the right side of (\ref{eq:condvarU}) are bounded above, and therefore $E(U_{t,k}^2)<\infty$ for all $k$ and $t$.
\end{proof}

\begin{proof}{\it [Theorem \ref{thm:staterg}]}.\\
	We start by representing the process $Z_t$ in an alternative form and by showing a stochastic order property of a particular beta distribution on the interval  $(0,1)$, which will be important steps to prove the theorem. First, note that for $Y\sim\mbox{Beta}(0,h)$, it can be checked that $Z=Y^2$ has density function given by
	\begin{eqnarray}\label{eq:betaZ}
		g_h(z)=\dfrac{\Gamma\left(\frac{1}{2h}\right)}{\sqrt{\pi}\Gamma\left(\frac{1/h-1}{2}\right)}z^{-1/2}(1-z)^{(1/h-3)/2}, \quad  z\in(0,1),
	\end{eqnarray}
	i.e., $Z$ follows a beta distribution on the interval $(0,1)$ with shape parameters $1/2$ and $(1/h-1)/2$. Denote by $G_h(\cdot)$ the cumulative distribution function associated with the density (\ref{eq:betaZ}), which depends on the parameter $h$, and define $\{U_t\}_{t\geq1}$ to be a sequence of independent and identically distributed uniform random variables on the interval (0,1). Then, the process $Z_t=Y_t^2$ can be expressed by
	\begin{eqnarray}\label{storep}
		Z_t=G^{-1}_{h_t}(U_t),\quad t\geq1.
	\end{eqnarray}	
	
	Let us now show a stochastic order result on the distribution associated with (\ref{eq:betaZ}). Let $Z'$ and $Z''$ be two beta random variables on the interval $(0,1)$ with density function assuming the form (\ref{eq:betaZ}) with associated parameters $h'$ and $h''$, respectively. We will show that, if $h'\leq h''$, then $Z'$ is stochastically smaller than $Z''$.
	
	Assume that $h'\leq h''$. The log-ratio of the densities of $Z'$ and $Z''$ is
	\begin{eqnarray*}
		r(z)=\log\left(\dfrac{g_{h'}(z)}{g_{h''}(z)}\right)=\left(\dfrac{1}{h'}-\dfrac{1}{h''}\right)\log(1-z)+K\left(\dfrac{1}{h'}\right)-K\left(\dfrac{1}{h''}\right),
	\end{eqnarray*}
	where we have defined $K(\eta)=\log\Gamma\left(\eta/2\right)-\log\Gamma\left((\eta-1)/2\right)$, for $\eta>0$. We have that $r(z)$ is concave function in $z$ (under the assumption $h'\leq h''$). Moreover, $K'(\eta)=dK(\eta)/d\eta=\{\psi(\eta/2)-\psi((\eta-1)/2)\}/2>0$, where the inequality follows from the fact that $\psi(\cdot)$ is a strictly increasing function. Therefore, $K(\eta)$ is strictly increasing in $\eta$ (since its derivative is always positive). Hence, 
	\begin{eqnarray}\label{eq:auxso}
		\lim_{z\rightarrow0+}r(z)=K\left(\dfrac{1}{h'}\right)-K\left(\dfrac{1}{h''}\right)\geq0
	\end{eqnarray}	
	since $1/h'\geq 1/h''$. With the concavity of $r(z)$ and inequality (\ref{eq:auxso}), Theorem 1 by \cite{yu2009} gives that $Z'$ is stochastically smaller than $Z''$ if $h'\leq h''$.
	
	We now have key preliminary results to proceed. We will show that there is a unique strictly stationarity solution to $\{(Z_t,h_t)\}_{t\geq1}$ by checking that Condition 3.1 from \cite{douwin2008} holds (the other two conditions from that paper follow straightforwardly in our case). Let $W_t=(Z_t,h_t)^\top$, for $t\geq1$. From (\ref{storep}), there is a function $F(\cdot,\cdot)$ such that $W_t=F(W_{t-1},U_t)$, where $\{U_t\}_{t\geq1}$ is a sequence of iid (uniform) random variables. Now, for $\omega=(z,h)^\top$ and $\epsilon>0$, define $\|\omega\|_\epsilon=|z|+\epsilon|h|$. We want to show that there is constant $c\in(0,1)$ such that
	\begin{eqnarray}\label{cond3.1}
		E\left(\|F(\omega,U_t)-F(\omega',U_t)\|_\epsilon\right)\leq c \|\omega-\omega'\|_\epsilon,
	\end{eqnarray}	
	for some $\epsilon>0$, where $\omega'=(z',h')^\top$. It follows that
	\begin{eqnarray}\label{proof:ineqaux}
		&&E\left(\|F(\omega,U_t)-F(\omega',U_t)\|_\epsilon\right)=\nonumber\\
		&&E\left(\left|G^{-1}_{\alpha_0+\alpha_1 z+\beta_1 h}(U_t)-G^{-1}_{\alpha_0+\alpha_1 z'+\beta_1 h'}(U_t)\right|\right)+\epsilon|\alpha_0+\alpha_1 z+\beta_1 h-(\alpha_0+\alpha_1 z'+\beta_1 h')|.
	\end{eqnarray}	
	
	We now use the stochastic order results in our preliminaries at the beginning of the proof to compute the expectation in (\ref{proof:ineqaux}).
	If $\alpha_0+\alpha_1 z+\beta_1 h\leq \alpha_0+\alpha_1 z'+\beta_1 h'$, the stochastic order result of the $\mbox{Beta}(0,h)$ distributions gives us that $G^{-1}_{\alpha_0+\alpha_1 z+\beta_1 h}(U_t)\geq G^{-1}_{\alpha_0+\alpha_1 z'+\beta_1 h'}(U_t)$, and therefore 
	\begin{eqnarray*}
		E\left(\left|G^{-1}_{\alpha_0+\alpha_1 z+\beta_1 h}(U_t)-G^{-1}_{\alpha_0+\alpha_1 z'+\beta_1 h'}(U_t)\right|\right)&=&E\left(G^{-1}_{\alpha_0+\alpha_1 z+\beta_1 h}(U_t)-G^{-1}_{\alpha_0+\alpha_1 z'+\beta_1 h'}(U_t)\right)\\
		&=&\alpha_0+\alpha_1 z+\beta_1 h-(\alpha_0+\alpha_1 z'+\beta_1 h').
	\end{eqnarray*}
	Otherwise, $E\left(\left|G^{-1}_{\alpha_0+\alpha_1 z+\beta_1 h}(U_t)-G^{-1}_{\alpha_0+\alpha_1 z'+\beta_1 h'}(U_t)\right|\right)=\alpha_0+\alpha_1 z'+\beta_1 h'-(\alpha_0+\alpha_1 z+\beta_1 h)$. Both cases can be expressed by 
	$$E\left(\left|G^{-1}_{\alpha_0+\alpha_1 z+\beta_1 h}(U_t)-G^{-1}_{\alpha_0+\alpha_1 z'+\beta_1 h'}(U_t)\right|\right)=|\alpha_0+\alpha_1 z+\beta_1 h-(\alpha_0+\alpha_1 z'+\beta_1 h')|.$$
	
	By using this in (\ref{proof:ineqaux}), we obtain that
	\begin{eqnarray*}
		E\left(\|F(\omega,U_t)-F(\omega',U_t)\|_\epsilon\right)&=&(1+\epsilon)|\alpha_1 (z-z')+\beta_1(h-h')|\leq \\
		&&(1+\epsilon)(\alpha_1 |z-z'|+\beta_1|h-h'|)\leq (1+\epsilon)\max(\alpha_1,\beta_1/\epsilon)\|\omega-\omega'\|_\epsilon.
	\end{eqnarray*}	
	
	By choosing $\epsilon=\beta_1/\alpha_1$, we have that $(1+\epsilon)\max(\alpha_1,\beta_1/\epsilon)=\alpha_1+\beta_1<1$, where the last inequality follows from model definition. Therefore, Condition 3.1 from \cite{douwin2008} holds (Inequality (\ref{cond3.1})) and we obtain the existence and uniqueness of the stationary solution to $\{(Z_t,h_t)\}_{\geq1}$, which gives us part (i). Moreover, the results from  \cite{douwin2008} also gives us that the process $\{(Z_t,h_t)\}_{\geq1}$ is ergodic when initiated with $(Z_0,h_0)\sim\Pi$, with $\Pi$ denoting the stationary distribution, which completes the proof of the theorem part (ii).
\end{proof}

\begin{proof}{\it [Theorem \ref{thm:AN}]}.\\ 
	We will show that the following three conditions are satisfied:\\
	\noindent	(C1) $\dfrac{1}{\sqrt{n}}\dfrac{\partial\ell(\boldsymbol\theta)}{\partial\boldsymbol\theta}\stackrel{d}{\longrightarrow}N({\bf 0},\boldsymbol\Sigma_S)$, $\boldsymbol\Sigma_S>0$;\\
	\noindent	(C2) $-\dfrac{1}{n}\dfrac{\partial^2\ell(\boldsymbol\theta)}{\partial\boldsymbol\theta\partial\boldsymbol\theta^\top}\stackrel{p}{\longrightarrow}\boldsymbol\Sigma_I$, $\boldsymbol\Sigma_I>0$;\\
	\noindent	(C3) $\max_{l,i,j}\sup_{\boldsymbol\theta\in N(\boldsymbol\theta_0)}\dfrac{1}{n}\left|\dfrac{\partial^3\ell(\boldsymbol\theta)}{\partial\theta_l\partial\theta_i\partial\theta_j}\right|\leq C_n$, where $N(\boldsymbol\theta_0)$ is a neighborhood of $\boldsymbol\theta_0$ and $0\leq C_n\stackrel{p}{\longrightarrow} c$, with $0<c<\infty$.
	
	Then, the result will follow from Lemma 3.1 by \cite{jenrah2004}. As a note, although the paper by these authors is about asymptotics for non-stationary GARCH models, their Lemma 3.1 holds in general. As stated by the authors in their Section 3, Conditions (C1), (C2), and (C3) are similar to conditions stated for general asymptotic theory literature; for instance, \citep{leh1999,basetal1976}. The difference is that Conditions (C1), (C2), and (C3) explicitly involves third order derivatives of the log-likelihood function and their lemma provides uniqueness in addition to existence of the maximum likelihood estimators. Also, we have slightly adapted the conditions from \cite{jenrah2004} since they define the log-likelihood function with a multiplicative factor of $-1/n$, with $n$ denoting the sample size. 
	
	Under the hypotheses of the present theorem and from Theorems \ref{thm:staterg} and \ref{thm:mds}, we have that $\{{\bf U}_t(\boldsymbol\theta)\}_{t\geq1}$ is a stationary ergodic martingale difference sequence with finite second moment. We can then apply the Lindeberg-L\'evy Theorem for martingales \citep{bil1961} to obtain the weak convergence in (C1), with $\boldsymbol\Sigma_S=\mbox{cov}({\bf U}_t(\boldsymbol\theta))$. The diagonal elements of $\mbox{cov}({\bf U}_t(\boldsymbol\theta))$ are provided in Equation (\ref{eq:condvarU}). The off-diagonal elements can also be computed explicitly. For $k\neq j$, it follows that 
	\begin{eqnarray}\label{eq:cov}
		\mbox{cov}(U_{t,k},U_{t,j})&=&\dfrac{1}{4}E\left[\left(-\log(1-Y_t^2)-\psi\left(\frac{1}{2h_t}\right)+\psi\left(\frac{1/h_t-1}{2}\right)\right)^2\dfrac{1}{h^4_t}
		\dfrac{\partial h_t}{\partial\theta_k}\dfrac{\partial h_t}{\partial\theta_j}\right]\nonumber\\
		&=&\dfrac{1}{4}E\left[\mbox{Var}\left(-\log(1-Y_t^2)\bigg|\mathcal F_{t-1}\right)\dfrac{1}{h^4_t}
		\dfrac{\partial h_t}{\partial\theta_k}\dfrac{\partial h_t}{\partial\theta_j}\right]\nonumber\\
		&=&\dfrac{1}{4}E\left[\left(\psi'\left(\frac{1/h_t-1}{2}\right)-\psi'\left(\frac{1}{2h_t}\right)\right)\dfrac{1}{h_t^4}\dfrac{\partial h_t}{\partial\theta_k}\dfrac{\partial h_t}{\partial\theta_j}\right].
	\end{eqnarray}	
	
	By using the expressions of the covariances and variances given in (\ref{eq:condvarU}) and (\ref{eq:cov}), respectively, we obtain that $\boldsymbol\Sigma_S=\boldsymbol\Sigma$, with $\boldsymbol\Sigma$ as defined in (\ref{eq:Sigma}). We now claim that $\boldsymbol\Sigma$ is positive definite (notation: $\boldsymbol\Sigma>0$). For a $1\times 3$ vector ${\bf u}\neq{\bf0}$, we have that ${\bf u}\left(\dfrac{\partial h_t}{\partial\boldsymbol\theta}\dfrac{\partial h_t}{\partial\boldsymbol\theta}^\top\right){\bf u}^\top={\bf u}\dfrac{\partial h_t}{\partial\boldsymbol\theta}\left({\bf u}\dfrac{\partial h_t}{\partial\boldsymbol\theta}\right)^\top=\left({\bf u}\dfrac{\partial h_t}{\partial\boldsymbol\theta}\right)^2>0$ with probability 1. Therefore, $\dfrac{\partial h_t}{\partial\boldsymbol\theta}\dfrac{\partial h_t}{\partial\boldsymbol\theta}^\top>0$. Note also that $\psi'\left(\frac{1/h_t-1}{2}\right)-\psi'\left(\frac{1}{2h_t}\right)>0$ since $\psi'(\cdot)$ is a strictly decreasing function. These arguments are sufficient to establish that the random matrix appearing in the expectation (\ref{eq:Sigma}) is positive definite, and hence its expectation is also positive definite.
	
	We now apply the Law of Large Numbers for stationary and ergodic sequences (for instance, see \cite{bre1992}) to obtain Condition (C2): 
	\begin{eqnarray*}
		\plim_{n\rightarrow\infty}-\dfrac{1}{n}\dfrac{\partial^2\ell(\boldsymbol\theta)}{\partial\boldsymbol\theta\partial\boldsymbol\theta^\top}=E\left(-\dfrac{\partial{\bf U}_t(\boldsymbol\theta)}{\partial\boldsymbol\theta^\top}\right)\equiv\boldsymbol\Sigma_I,
	\end{eqnarray*}
	where
	\begin{eqnarray*}
		\dfrac{\partial{\bf U}_t(\boldsymbol\theta)}{\partial\boldsymbol\theta^\top}&=&-\left\{\psi'\left(\frac{1/h_t-1}{2}\right)-\psi'\left(\frac{1}{2h_t}\right)\right\}\dfrac{1}{4h_t^2}\dfrac{\partial h_t}{\partial\boldsymbol\theta}\dfrac{\partial h_t}{\partial\boldsymbol\theta}^\top+\\
		&&\left\{-\log(1-Y_t^2)-\psi\left(\frac{1}{2h_t}\right)+\psi\left(\frac{1/h_t-1}{2}\right)\right\}\left\{-\dfrac{1}{h_t^3}\dfrac{\partial h_t}{\partial\boldsymbol\theta}\dfrac{\partial h_t}{\partial\boldsymbol\theta}^\top+\dfrac{1}{2h_t^2}\dfrac{\partial^2 h_t}{\partial\boldsymbol\theta\partial\boldsymbol\theta^\top}\right\}.
	\end{eqnarray*}
	Hence, it can be checked that $\boldsymbol\Sigma_I\equiv E\left(-\partial{\bf U}_t(\boldsymbol\theta)/\partial\boldsymbol\theta^\top\right)=\boldsymbol\Sigma>0$.
	
	The last step consists of showing that Condition (C3) is satisfied. We will consider the case involving three derivatives of the conditional log-likelihood function with respect to the parameter $\alpha_1$. The remaining  cases can be argued in a similar fashion and therefore are omitted. We have that
	\begin{eqnarray*}
		\dfrac{\partial\ell(\boldsymbol\theta)}{\partial\alpha_1^3}&=&\sum_{t=2}^n\bigg\{\left[\psi''\left(\frac{1/h_t-1}{2}\right)-\psi''\left(\frac{1}{2h_t}\right)\right]\dfrac{1}{8h_t^6}\left(\dfrac{\partial h_t}{\partial\alpha_1}\right)^3+\\
		&&\left[\psi'\left(\frac{1/h_t-1}{2}\right)-\psi'\left(\frac{1}{2h_t}\right)\right]\dfrac{1}{h_t^5}\left(\dfrac{\partial h_t}{\partial\alpha_1}\right)^3+ \\ 	 	
		&& \left[-\log(1-Y_t^2)-\psi\left(\frac{1}{2h_t^2}\right)+\psi\left(\frac{1/h_t-1}{2}\right)\right]\dfrac{3}{h_t^4}\left(\dfrac{\partial h_t}{\partial\alpha_1}\right)^3 -	\\
		&&\left[\psi\left(\frac{1/h_t-1}{2}\right)-\psi\left(\frac{1}{2h_t}\right)\right]\dfrac{1}{h_t^3}\left(\dfrac{\partial h_t}{\partial\alpha_1}\right)^2 	 		
		\bigg\}.
	\end{eqnarray*}	
	
	By applying recursion $t$ times, we can express $h_t$ as $h_t=\alpha_0\dfrac{1-\beta_1^{t-1}}{1-\beta_1}+\alpha_1\displaystyle\sum_{j=1}^{t-1}\beta_1^{j-1}Y_{t-j}^2+\beta_1^{t-1}h_1$. Hence, we get $h_t\geq \alpha_1\displaystyle\sum_{j=1}^{t-1}\beta_1^{j-1}Y_{t-j}^2=\alpha_1 \dfrac{\partial h_t}{\partial\alpha_1}>0$. Therefore, $0\leq \left( \dfrac{\partial h_t}{\partial\alpha_1}/h_t\right)^k\leq\alpha_1^k$, for all $k\geq1$.
	Moreover, $\alpha_0\leq h_t\leq \alpha_0+\alpha_1+\beta_1<1$ for all $t$, and then $\dfrac{1}{\alpha_0+\alpha_1+\beta_1}\leq \dfrac{1}{h_t}\leq \dfrac{1}{\alpha_0}$, which is bounded away from 1. These bounds for $1/h_t$ and the continuity of the functions $\psi(\cdot)$, $\psi'(\cdot)$, and $\psi''(\cdot)$ imply that $\left|\psi\left(\frac{1/h_t-1}{2}\right)-\psi\left(\frac{1}{2h_t}\right)\right|$, $\left|\psi'\left(\frac{1/h_t-1}{2}\right)-\psi'\left(\frac{1}{2h_t}\right)\right|$, and $\left|\psi''\left(\frac{1/h_t-1}{2}\right)-\psi''\left(\frac{1}{2h_t}\right)\right|$ are bounded above by a finite constant. Using the above arguments, there are constants $c_1\in(0,\infty)$ and $c_2\in(0,\infty)$ such that
	\begin{eqnarray*}
		\dfrac{1}{n}\left|\dfrac{\partial\ell(\boldsymbol\theta)}{\partial\alpha_1^3}\right|&\leq& c_1\dfrac{1}{n}\sum_{t=2}^n-\log(1-Y_t^2)+c_2\dfrac{n-1}{n}\\
		&\stackrel{a.s.}{\longrightarrow}&c_1E\left(\psi'\left(\frac{1}{2h_t}\right)-\psi'\left(\frac{1/h_t-1}{2}\right)\right)+c_2<\infty,
	\end{eqnarray*}	
	where we have applied the Law of Large Number for stationary and ergodic sequences to get the almost sure convergence above. Since Conditions (C1), (C2), and (C3) are satisfied, Lemma 3.1 by \cite{jenrah2004} gives us the existence, uniqueness, and consistency of $\widehat{\boldsymbol\theta}$ stated in (i) and (ii), and its asymptotic normality with null mean vector and covariance matrix $\boldsymbol\Omega_I^{-1}\boldsymbol\Omega_S\boldsymbol\Omega_I^{-1}=\boldsymbol\Sigma^{-1}\boldsymbol\Sigma\boldsymbol\Sigma^{-1}=\boldsymbol\Sigma^{-1}$, as stated in (iii).
\end{proof}

\begin{proof}{\it [Theorem \ref{thm:apporx_ell}]}.\\  
	Without loss of generality, assume that the set $V$ assumes the form $V=(\underline\alpha_0,\overline\alpha_0)\times(\underline\alpha_1,\overline\alpha_1)\times(\underline\beta_1,\overline\beta_1)$, which contains the true parameter vector, with $0<\underline\alpha_0,\underline\alpha_1,\underline\beta_1$, $\underline\alpha_0<\overline\alpha_0$, $\underline\alpha_1<\overline\alpha_1$, $\underline\beta_1<\overline\beta_1$, and $\overline\alpha_0,\overline\alpha_1,\overline\beta_1<1$. Now, by applying recursion of $h_t$ $t$ times (like done in proof of Theorem \ref{thm:AN}), we obtain that $|h_t-\widetilde h_t|=\beta_1^{t-1}|h_1-\widetilde h_1|$, and hence 
	\begin{eqnarray}\label{ineq:tildeh}
		\sup_{\boldsymbol\theta\in V}|h_t-\widetilde h_t|\leq 2{\overline\beta_1}^{t-1}.
	\end{eqnarray}	
	Also,
	\begin{eqnarray}\label{ineq:loglikdiffs}
		\left|\ell(\boldsymbol\theta)-\widetilde\ell(\boldsymbol\theta)\right|\leq\sum_{t=2}^n\dfrac{-\log\left(1-Y_t^2\right)}{2}\left|\dfrac{1}{\widetilde h_t}-\dfrac{1}{h_t}\right|+\sum_{t=2}^n\left|\log\Gamma\left(\dfrac{1}{2\widetilde h_t}\right)-\log\Gamma\left(\dfrac{1}{2 h_t}\right)\right|\nonumber\\
		+\sum_{t=2}^n\left|\log\Gamma\left(\dfrac{1/\widetilde h_t-1}{2}\right)-\log\Gamma\left(\dfrac{1/h_t-1}{2}\right)\right|.
	\end{eqnarray}	
	
	The next step is to find upper bounds for the terms involved on the right side of the inequality (\ref{ineq:loglikdiffs}). By using that $h_t,\widetilde h_t\geq\alpha_0$ and (\ref{ineq:tildeh}), if follows that 
	\begin{eqnarray}\label{hdiffs1}
		\sup_{\boldsymbol\theta\in V}\left|\dfrac{1}{\widetilde h_t}-\dfrac{1}{h_t}\right|=\sup_{\boldsymbol\theta\in V}\dfrac{\left|h_t-\widetilde h_t\right|}{\widetilde h_th_t}\leq\sup_{\boldsymbol\theta\in V}\dfrac{1}{\alpha_0^2}\left|h_t-\widetilde h_t\right|\leq\dfrac{2{\overline\beta_1}^{t-1}}{\underline\alpha_0^2}\equiv 2d_1{\overline\beta_1}^{t-1}.
	\end{eqnarray}
	
	Now, for $0<x<y$, we use the Mean Value Theorem to obtain that $\log\Gamma(x)-\log\Gamma(y)=(x-y)\psi(z)$, where $\psi(\cdot)$ is the digamma function and $z\in(x,y)$. This and the inequality $\log x-\dfrac{1}{x}<\psi(x)<\log x-\dfrac{1}{2x}$ (for instance, see \cite{alz1997}) give us that $|\log\Gamma(x)-\log\Gamma(y)|\leq|x-y|\max\left(|\log z-\frac{1}{z}|,|\log z-\frac{1}{2z}|\right)$. For $z\geq\alpha_0$, it can be checked that $\left|\log (2z)^{-1}-\frac{1}{(1/2z)}\right|\leq\log2+2-\log\alpha_0$ and $\left|\log (2z)^{-1}-\frac{1}{(1/z)}\right|\leq\log2+1-\log\alpha_0$. Using these results, the fact that $h_t,\widetilde h_t\geq\alpha_0$, and (\ref{ineq:tildeh}), we get that
	\begin{eqnarray}\label{hdiffs2}
		\sup_{\boldsymbol\theta\in V}\left|\log\Gamma\left(\dfrac{1}{2\widetilde h_t}\right)-\log\Gamma\left(\dfrac{1}{2 h_t}\right)\right|&\leq&\sup_{\boldsymbol\theta\in V}\dfrac{\log2+2-\log\alpha_0}{2}\left|\dfrac{1}{\widetilde h_t}-\dfrac{1}{h_t}\right|\nonumber\\
		&\leq&
		\dfrac{\log2+2-\log\underline\alpha_0}{2\underline\alpha_0^2}2{\overline\beta_1}^{t-1}\equiv d_2{\overline\beta_1}^{t-1}.
	\end{eqnarray}	
	
	Similarly, it can be argued that there exists constant $d_3\in(0,\infty)$ such that
	\begin{eqnarray}\label{hdiffs3}
		\sup_{\boldsymbol\theta\in V}\left|\log\Gamma\left(\dfrac{1/\widetilde h_t-1}{2}\right)-\log\Gamma\left(\dfrac{1/h_t-1}{2}\right)\right|
		\leq
		d_3{\overline\beta_1}^{t-1}.
	\end{eqnarray}	
	
	From (\ref{ineq:loglikdiffs})-(\ref{hdiffs3}), it follows that 
	\begin{eqnarray}\label{ineq:loglikdiffs2}
		\sup_{\boldsymbol\theta\in V}	\left|\dfrac{1}{n}\ell(\boldsymbol\theta)-\dfrac{1}{n}\widetilde\ell(\boldsymbol\theta)\right|\leq\dfrac{1}{n}\sum_{t=2}^n\left(-\log\left(1-Y_t^2\right)d_1+d_2+d_3\right){\overline\beta_1}^{t-1}.
	\end{eqnarray}	
	
	For each $\epsilon>0$, we obtain from (\ref{ineq:loglikdiffs2}) that 
	\begin{eqnarray*}
		\limsup_{n\rightarrow\infty}	P\left(\left|\dfrac{1}{n}\ell(\boldsymbol\theta)-\dfrac{1}{n}\widetilde\ell(\boldsymbol\theta)\right|>\epsilon\right)\leq
		\limsup_{n\rightarrow\infty}	P\left(\dfrac{1}{n}\sum_{t=2}^n\left(-\log\left(1-Y_t^2\right)d_1+d_2+d_3\right){\overline\beta_1}^{t-1}>\epsilon\right) \leq\\
		\limsup_{n\rightarrow\infty}\dfrac{d_1E(-\log(1-Y_1^2))+d_2+d_3}{n\epsilon}\sum_{t=2}^n{\overline\beta_1}^{t-1}=\dfrac{d_1E(-\log(1-Y_1^2))+d_2+d_3}{\epsilon(1-\beta_1)}\lim_{n\rightarrow\infty}\dfrac{\overline\beta_1-{\overline\beta_1}^{n}}{n}=0,
	\end{eqnarray*}	
	which gives us the desired result, where we have used the Markov inequality to obtain the second inequality above and the facts that $Y_t^2$ is strictly stationary under the conditions of Theorem \ref{thm:staterg} and that $E(-\log(1-Y_1^2))<\infty$.
\end{proof}

\noindent {\bf Conflicts of interest:} None declared. \\

\noindent {\bf Data availability:} The data are publicly available from \url{https://finance.yahoo.com}.

\end{document}